\documentclass[a4paper,11pt]{article}

\usepackage{jheppub} 
\usepackage[T1]{fontenc} 
\usepackage{tikz, pgf}
\usetikzlibrary{shapes.misc}
\usetikzlibrary{shapes}
\usetikzlibrary{fit}
\usetikzlibrary{arrows}
\def\rgnote#1{{\color{magenta} #1}}
\def\arnote#1{{\color{red} #1}}
\def\mvnote#1{{\color{black} #1}}
\def\ssnote#1{{{\color{OrangeRed}#1}}}
\def\nn{\nonumber}

\def\iimg{ i}

\def\iifty{\iimg \infty}

\def\piimg{\pi \iimg}
\def\inc{\gamma}
\def\exc{\Delta}
\def\mv{s}
\def\inmv{ w}
\def\lpvr{\mathcal{R}}
\def\trvr{ R}
\def\lpvp{\mathcal{P}}
\def\lpvq{\mathcal{Q}}

\def\lpmv{v}

\newcommand{\be}{\begin{eqnarray}}
\newcommand{\ee}{\end{eqnarray}}

\newcommand{\bi}{\begin{itemize}}
\newcommand{\ei}{\end{itemize}}

\newcommand{\bse}{\begin{subequations}}
\newcommand{\ese}{\end{subequations}}

\newcommand{\non}{\nonumber}

\title{Exploring Perturbative Conformal Field Theory in Mellin space}

\author[a]{Amin A. Nizami,}
\author[b]{Arnab Rudra,}
\author[c,d]{Sourav Sarkar}
\author[a,e]{and Mritunjay Verma}

\vspace*{1cm}

\affiliation[a] {\it International Centre for Theoretical Sciences, TIFR, Hesaraghatta, Hubli, Bengaluru-560089, India }
\affiliation[b] {\it  Center for Quantum Mathematics and Physics (QMAP), Department of Physics, University of California, Davis, 1 Shields Ave, Davis, CA 95616, United States}
\affiliation[c]{\it Institut f{\"u}r Mathematik und Institut f{\"u}r Physik, Humboldt-Universit{\"a}t zu Berlin, IRIS-Adlershof, Zum Gro{\ss}en Windkanal 6, 12489 Berlin, Germany}
\affiliation[d]{\it Max-Planck-Institut f{\"u}r Gravitationsphysik, Albert-Einstein-Institut, Am M{\"u}hlenberg 1, 14476 Potsdam, Germany}
\affiliation[e]{\it Harish-Chandra Research Institute, Chhatnag Road, Jhunsi, Allahabad-211019, India }

\vspace{1cm}

\emailAdd{amin@icts.res.in}

\emailAdd{rudra@ucdavis.edu}

\emailAdd{sarkar@physik.hu-berlin.de}

\emailAdd{mritunjayverma@hri.res.in}

\abstract{We explore the Mellin representation of correlation functions in conformal field theories in the weak coupling regime. We provide a complete proof for a set of Feynman rules to write the Mellin amplitude for a general tree level Feynman diagram involving only scalar operators.  We find a factorised form involving beta functions associated to the propagators, similar to tree level Feynman rules in momentum space for ordinary QFTs. We also briefly consider the case where a generic scalar perturbation of the free CFT breaks conformal invariance. Mellin space still has some utility and one can consider non-conformal Mellin representations. In this context, we find that the beta function corresponding to conformal propagator uplifts to a hypergeometric function.}

\preprint{\parbox{3cm}{HRI/ST/1602\\ HU-EP-16/23\\ HU-MATH-2016/14 \\ICTS/2016/4}}

\begin{document} 
\maketitle
\flushbottom

\section{Introduction}
It has been realised in the last few years, beginning  with the pioneering work of Mack \cite{Mack1, Mack2} (see also \cite{symanzik}), that Mellin space provides the natural setting for the study of Conformal Field theories (CFTs). The Mellin transform of a CFT correlator is a meromorphic function in the Mellin variables. In particular, for a four point function, the isolated simple poles locate the conformal twists of the operators in the spectrum whereas the residues at these poles contain information about the 3-point couplings. Thus the CFT data (operator dimensions and OPE coefficients) is at once made manifest in the Mellin space representation. Mellin amplitudes are also conformally invariant making conformal symmetry manifest in Mellin space.

\vspace*{0.06in}Usually in quantum field theory, we Fourier transform the position space correlators to write Feynman rules in momentum space. The important advantage in doing so is that translation invariance leads to momentum conservation and the position space integrals are reduced to simple products in momentum space at tree level. In momentum space, conformal transformations have a non-linear action  and as a result the conventional way of doing perturbative QFT in momentum space is not so advantageous for CFTs. 

\vspace*{.06in}Various important features of QFT such as locality, causality and unitarity can be understood in terms of the analytic properties of momentum space amplitudes. The isolated poles of the momentum space propagator correspond to single-particle states and the branch cuts on the real axis give the multi-particle states (K{\"a}hlen Lehmann spectral representation) and the amplitudes factorise on residues at the poles to lower point amplitudes. In a CFT, we do not have single particle states characterised by the masses since mass is a dimensionful parameter. Hence the propagators in momentum space have branch cuts extending to the origin. In the radial quantization of CFT, the dilatation operator acts as the Hamiltonian. The eigenvalues of this operator are discrete for $d>2$. This discrete set of operators appear in the operator product expansion (OPE) as the exchanged primaries and descendants in an interacting CFT (in $d>2$) . So it is desirable to have a representation for correlation functions in CFTs that makes this discrete spectrum manifest. As shown by Mack, it turns out that Mellin space provides such a representation. 

\vspace*{0.06in}The analogy of the Mellin space CFT correlators with scattering amplitudes is also striking. This has been explored in the context of the AdS/CFT correspondence. Following Mack, the application of the Mellin representation of conformal correlation functions was explored at strong coupling for large $N$ CFTs using tree level Witten diagrams in $AdS$ \cite{pened,suvrat,paulos2, dhriti, costa,perlmutter, kaplan}. While at tree level, there seem to be a set of Feynman rules to write the Mellin amplitudes, the loop level seems to be significantly more involved. In the flat space limit of $AdS/CFT$, a relation between the bulk scattering amplitude and the CFT Mellin amplitudes was also suggested in \cite{pened,suvrat} and later put on a firm footing in \cite{analytic, unitary,kaplan,factorization} (see also \cite{bootstrap1}). To be precise, the flat space S-matrix is expressed as an integral transform of the CFT Mellin amplitude and the Mellin variables, in the flat space limit, turn into flat space kinematic invariants (the Mandelstam variables). This scheme also relates the S-Matrix program in QFTs to the Bootstrap program in CFTs. Our work, however, has a different focus and does not use AdS/CFT. We consider {\it weakly} coupled CFTs and attempt to formulate Feynman rules in Mellin space for perturbative field theory computations.

\vspace*{.06in}The Mellin representation for tensor operators and the factorization of Mellin amplitudes was studied in \cite{factorization}. The Mellin representation has also been explored in the context of minimal model CFTs in \cite{lowe} and for open string amplitudes in \cite{stieberger}. It was explored in the weak coupling regime in \cite{paulos1,dhriti2} in the context of SYM and has also been used to calculate corrections beyond the planar limit to the 4-point function of a primary in $\mathcal{N}=4$ SYM in \cite{vasco2}. For some more applications in the context of $\mathcal{N}=4$ SYM, see also \cite{alday, zhibo, zhibo2,zhibo3,Alday}. Feynman rules for tree level diagrams in the Mellin space were stated in \cite{paulos1,dhriti2} after considering a few examples. However a proof of these rules for a general tree level Feynman diagram was not provided. 

\vspace*{0.06in}The goal of our note is to further explore the suitability of the Mellin representation for studying perturbative CFTs. We consider an exactly marginal perturbation around a free CFT and investigate whether it is possible to obtain a set of Feynman rules that can be used to calculated Mellin amplitudes. For simplicity, we restrict to scalar operators throughout the paper. We present a complete derivation of the Feynman rules associated to tree level amplitudes in complete generality. For this purpose, we also develop a diagrammatic algorithm to write down the Mellin amplitude for any Feynman diagram (upto arbitrary loop order) as an integral over Schwinger parameters corresponding to the internal propagators in the diagram. We further relax the conformality of the integrals, we consider, to study Mellin amplitudes in free CFTs with a generic perturbation. It turns out that when we consider integrals that enjoy a scale covariance only (as opposed to the full conformal covariance) the corresponding ``Mellin amplitudes'' can be interpreted as ``off-shell'' quantities that reduce to the ``on-shell'' conformal Mellin amplitudes under an LSZ like prescription.  

\vspace*{0.06in}For application to many well-known conformal field theories (say, $\mathcal{N}=4$ Super-Yang Mills), we need to extend these rules to tensor and fermionic operators as well. We leave this along with the task of obtaining Feynman rules for loop amplitudes to future work.

\vspace*{.06in}The plan of this paper is as follows. In section \ref{sec:mellin}, we give a quick review of the Mellin amplitude for conformal field theories. In section \ref{TL}, we consider some simple tree level Feynman diagrams involving only scalar fields and derive their Mellin amplitude. This is to introduce the general strategy that we follow for deriving the Mellin amplitude of a general tree level Feynman diagram. In section \ref{sec:diag}, we provide a general derivation for the Feynman rules for tree level diagrams. For this, we develop an algorithmic method for writing down the Mellin amplitude for an arbitrary Feynman diagram (tree as well as loops) as an integral over the Schwinger parameters for the internal propagators. In section \ref{mellin_loop}, we consider the Mellin amplitudes for loop diagrams involving scalar fields. Even though we can write an integral expression for the Mellin amplitude for such diagrams, we have not been able to perform these integrals so as to obtain a set of Feynman rules. In section \ref{OA}, we extend the notion of Mellin amplitude to include generic scalar deformations of a free CFT which may break conformal invariance. In particular, we consider a tree level diagram with a single internal line (involving scalar fields) in Mellin space for such theories. The appendices elaborate on our notations and conventions, contain some properties of the Mellin transform and a few useful identites. Many elaborate details of the calculations are also relegated to the appendices.

\vspace*{.06in}Throughout the draft, the space-time Lorentz indices will be suppressed. We shall use the indices $\{i,j,\cdots\}$ for external vertices and the indices $\{a,b,\cdots\}$ for internal vertices. For convenience, we shall use the upstair indices for denoting the external vertices and the lower indices for denoting the internal vertices. This turns out to be useful for us mainly because of the fact that our analysis does not depend on how many external legs are attached to a given internal vertex. This will become clear when we consider explicit calculations. More details on the notations and convention can be found in the appendix \ref{sec:notation}.

\section{Mellin Amplitude}
\label{sec:mellin}
The Mellin amplitude for an arbitrary $n$-point function is defined by the Mellin transformation of the position space correlation function \cite{Mack1, Mack2}
\begin{equation}
A\left(\{x^{i}\}\right)=
\prod\limits_{1\le i<j\le n}\left(
\int_{-\iifty}^{\iifty}
\frac{d\mv^{ij}}{2\pi i}\ \Gamma\left(\mv^{ij}\right)\left(x^{i}-x^{j}\right)^{-2\mv^{ij}}
\right)
\prod\limits_{i=1}^{n}\delta\left(\Delta^i-\sum\limits_{j=1}^n\mv^{ij}\right)
M\left(\{\mv^{ij}\}\right)
\label{Mellin1}
\end{equation} 
Here $\mv^{ij}$ are the Mellin variables and $M\left(\{\mv^{ij}\}\right)$ is defined to be the Mellin amplitude. The variable $\Delta^i$ is the scaling dimension of the operator inserted at $x^i$. One strips $M\left(\{s^{ij}\}\right)$ of the factors of $\Gamma(s^{ij})$ for convenience. This turns out to be particularly useful for large $N$ gauge theories (in the context of the $AdS/CFT$ correspondence) where these Gamma functions account for the poles corresponding to the multi-particle states whereas $M\left(\{\mv^{ij}\}\right)$ accounts for poles corresponding to the single particle states.

\vspace*{.06in}The following are some important points to be noted:
\begin{enumerate}
\item The delta function constraints in the definition of Mellin amplitude \eqref{Mellin1} ensure the covariance of $A(\{x_i\})$ under conformal transformations. More precisely, under inversion  
\begin{eqnarray}
\left(x^{i}-x^{j}\right)^{2} \rightarrow \frac{\left(x^{i}-x^{j}\right)^{2}}{(x^{i})^{2}(x^{j})^{2}}
\non
\end{eqnarray}
The correlation function $A\left(\{x^{i}\}\right)$ transforms as 
\begin{eqnarray}
A\left(\{x^{i}\}\right)\rightarrow \left[\prod_{i=1}^{n}(x^{i})^{-2\Delta^{i}}\right] \ A\left(\{x^{i}\}\right)
\non
\end{eqnarray}
The delta function constraints $\sum\limits_{j\not=i}\mv^{ij}=\Delta^{i}$
ensure that both sides of \eqref{Mellin1} transform in the same way. 

\item The Mellin amplitude $M(\{s^{ij}\})$ is manifestly conformally invariant. The conformal transformations act on the position space variables $x^i$. The $x^i$ dependence of the expression \eqref{Mellin1} and the delta functions imposing constraints on the Mellin variables ensure that $A(\{x^i\})$ is conformally covariant.

\item The Mellin variables $\mv^{ij}$ are symmetric in $i$ and $j$. So the number of Mellin variables $s^{ij}$ is $n(n-1)/2$. However, due to the $n$ delta function constraints, the number of independent Mellin variables is only $\frac{n(n-3)}{2}$. This is also the number of independent  cross-ratios for $n$ points and the number of Mandelstam invariants for an $n-$point scattering amplitude. 

\item The delta function constraints can be solved in terms of the ``dual Mellin momenta'' \cite{Mack1} with $\mv^{ij}=k^{i}\cdot k^{j}$ and $(k^i)^2=- \Delta^i$ and overall Mellin momentum conservation $\sum_{i}k^{i}=0$. These are fictitious momenta associated with each $x^i$. We refer to $(k^i)^2=- \Delta^i$ as the ``on-shell'' condition for Mellin momenta.

\item Using the dual Mellin momenta mentioned above, one can define ``dual Mandelstam variables'' and express the Mellin amplitude in terms of these. To see an explicit example \cite{vasco}, we consider a 4-point function and define the Mandelstam variables $s$ and $t$ as 
\begin{eqnarray}
\nonumber s\equiv -(p^{1}+p^{2})^{2}=\Delta^{1}+\Delta^{2}-2s^{12}\;\;\;\;\;\;\;\;,\;\;\;\;\;\;t\equiv -(p^{1}+p^{3})^{2}=\Delta^{1}+\Delta^{3}-2s^{13} 
\nonumber
\end{eqnarray}
In terms of these Mandelstam variables, we can express the 4-point amplitude as
\begin{eqnarray}
&& A(x^i)=\left[\prod\limits_{i<j}(x^{ij})^{-2\Delta^{ij}}\right] A(u,v) 
\ee
\mvnote{where,
\be
\Delta^{14}&=&-\frac{\Delta^{2}+\Delta^{3}}{2}\qquad, \;\;\;\;\;\;\;\;  \Delta^{24}=\frac{\Delta^{2}+\Delta^{4}}{2}\qquad, \;\;\;\;\;\;\;\; \Delta^{34}=\frac{\Delta^{3}+\Delta^{4}}{2}  \nonumber\\
 \Delta^{12}&=&\frac{\Delta^{1}+\Delta^{2}}{2} \qquad\quad, \;\;\;\;\;\;\;\; \Delta^{13}=\frac{\Delta^{1}+\Delta^{3}}{2} \qquad,\;\;\;\;\;\;\;\;   \Delta^{23}=-\frac{\Delta^{1}+\Delta^{4}}{2}\non
  \end{eqnarray}}
$u$ and $v$ are the usual 4-point cross ratios and $A(u,v)$ is the inverse Mellin transform with respect to the above Mandelstam variables 
\begin{eqnarray}
\nonumber {A}(u,v)&=&\int_{-i\infty}^{i\infty}\frac{dt}{2\pi i}\int_{-i\infty}^{i\infty}\frac{ds}{2\pi i}\;M(s,t)u^{\frac{s}{2}}v^{-\frac{s+t}{2}}\Gamma\left[\frac{\Delta^{1}+\Delta^{2}-s}{2}\right]\Gamma\left[\frac{\Delta^{1}+\Delta^{3}-t}{2}\right]\\
&&\hspace*{-.25in}\Gamma\left[\frac{\Delta^{3}+\Delta^{4}-s}{2}\right]\Gamma\left[\frac{\Delta^{2}+\Delta^{4}-t}{2}\right]\Gamma\left[\frac{s+t-\Delta^{2}-\Delta^{3}}{2}\right]\Gamma\left[\frac{s+t-\Delta^{1}-\Delta^{4}}{2}\right] 
   \non        
\end{eqnarray}
The above equation illustrates the fact that the position space correlator can be expressed as the inverse Mellin transform of the Mellin amplitude and that the kinematical variables in the Mellin space are analogous to the Mandelstam variables. 

\end{enumerate}

\section{  Some Examples of Tree Diagrams}                    
 \label{TL}
In this section we consider a few simple tree level examples which will illustrate the general strategy we shall follow for deriving the Mellin amplitude of Feynman diagrams involving only scalar fields. Specifically, we shall be looking at the contact interaction diagram, the  diagram with one internal propagator and the diagram with two internal propagators. The Mellin amplitude for the contact interaction diagram and one propagator diagram were presented in \cite{paulos1}. We begin with these examples for pedagogy and completeness of our presentation. 
\subsection{Contact Interaction}
\label{T1}

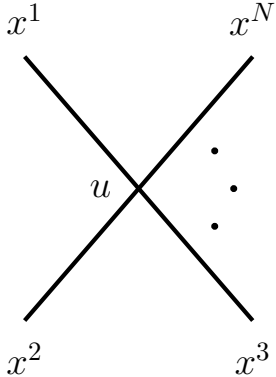
\begin{figure}[h]
\begin{center}
\begin{tikzpicture}[scale=.5]
\draw [-] [ultra thick](-3,-3.5) -- (0,0);
\draw [-] [ultra thick](3,-3.5) -- (0,0);
\draw [-] [ultra thick](3,3.5) -- (0,0);
\draw [-] [ultra thick](-3,3.5) -- (0,0);
\draw (-1,0) node{\Large $u$};
\draw (-3,-4.5) node{\Large $x^2$};
\draw (3,-4.5) node{\Large $x^3$};
\draw (3,4.5) node{\Large $x^N$};
\draw (-3,4.5) node{\Large $x^1$};
\begin{scope}[shift={(1,0)}] 
\filldraw [ultra thick] (1,1) circle (1pt);
\filldraw [ultra thick] (1.5,0) circle (1pt);
\filldraw [ultra thick] (1,-1) circle (1pt);
\end{scope}
\end{tikzpicture}
\end{center}
\caption{Contact Interaction Diagram}
\label{T1Vdiag1}
\end{figure}

The position space Feynman diagram for the contact interaction is shown in Figure \ref{T1Vdiag1}. In this diagram, $N$ external lines are meeting at the vertex $u$. We denote the scaling dimension of the field correponding to the external vertex $x^{i}$ by $\Delta^{i}$.  As mentioned earlier, we choose to place the index upstairs to keep the notation compact when we discuss more complicated Feynman diagrams. 

The position space correlation function corresponding to the contact interaction is given by
\begin{eqnarray}
I=\int \frac{d^{D}u}{2(2\pi)^{D/2}} \left[\prod\limits_{i=1}^{N} (x^{i}-u)^{-2\Delta^{i}}\Gamma(\Delta^i) \right]                \label{T1Veqn1}
\end{eqnarray}  
The factors of $\Gamma(\Delta^i)$  and $\pi$ have been included for the sake of convenience\footnote{In the rest of the draft, we denote the measure as
\begin{eqnarray}
\frac{d^Du}{2(2\pi)^{D/2}}\equiv \mathcal{D}u
\non\label{measure}
\end{eqnarray}
} later on. We follow these conventions throughout the draft.   

\vspace*{.06in}The expression is covariant under conformal transformations provided we impose the following `conformality condition' on the conformal dimensions
\begin{eqnarray}
\sum_{i=1}^N \Delta_i=D
\label{Conformality conditions}
\end{eqnarray}
\vspace*{.07in}We now introduce a Schwinger parameter for each propagator via the identity \begin{equation}
\frac{1}{(x-y)^{2\Delta}}=\frac{1}{\Gamma(\Delta)}\int_{0}^{\infty}d\alpha\;\alpha^{\Delta-1}\exp{\left[-\alpha(x-y)^{2}\right]}      \label{T1Veqn2}     
\end{equation}
\vspace*{.07in}Using this identity in \eqref{T1Veqn1} gives,
\begin{equation}
I=\prod_{i=1}^{N}\left[\int_0^\infty d\alpha^{i}(\alpha^{i})^{\Delta^{i}-1}\right]\int \frac{d^{D}u}{2(2\pi)^{D/2}}\exp{\left[-\left(\sum_{i=1}^{N}\alpha^{i}(x^{i}-u)^{2}\right)\right]}       \non            
\label{T1Veqn3}
\end{equation}
The factors of $\Gamma(\Delta^i)$ present in \eqref{T1Veqn1} are cancelled by the corresponding factors in \eqref{T1Veqn2}. 

\vspace*{.06in}Performing the Gaussian integration over $u$, we obtain
\begin{eqnarray}
I=\frac{1}{2}\prod_{i=1}^{N}\int_{0}^{\infty}d\alpha^{i}\left(\frac{(\alpha^{i})^{\Delta^{i}-1}}{(\sum\limits_{i} \alpha^{i})^{\frac{D}{2}}}\right)\exp{\left[-\frac{1}{\sum\limits_{i}\alpha^{i}}\left(\sum_{j}\sum_{i<j}\alpha^{i}\alpha^{j}(x^{ij})^{2}\right)\right]}                              \label{T1Veqn5}    
\end{eqnarray}
where, $x^{ij}\equiv x^{i}-x^{j}$.

\vspace*{.06in} We shall now render \eqref{T1Veqn5} particularly suitable for imposing the conformality conditions \eqref{Conformality conditions}. For this, we insert the following partition of unity in \eqref{T1Veqn5}
\be
1=\int_0^\infty dv \ \delta\left(v-\sum_{i}\alpha^i\right)
\non
\ee
We then rescale the Schwinger parameters, $\alpha^{i}\rightarrow \sqrt{v}\ \alpha^{i}$ and perform the integration over the auxiliary variable $v$ using delta function. The end result is
\begin{eqnarray}
I=\prod_{i=1}^{N}\left[\int_{0}^{\infty}d\alpha^{i}(\alpha^{i})^{\Delta^{i}-1}\right]\left(\sum_{i} \alpha^{i}\right)^{\sum\limits_{i}\Delta^i-D}\exp{\left(-\sum_{j}\sum_{i<j}\alpha^{i}\alpha^{j}(x^{ij})^{2}\right)}                             
 \label{T1Veqn6}    
\end{eqnarray}
\vspace*{.04in} We now impose the conformality condition \eqref{Conformality conditions} on equation \eqref{T1Veqn6} to obtain
\begin{eqnarray}
I=\prod_{i=1}^{N}\left[\int_{0}^{\infty}d\alpha^{i}(\alpha^{i})^{\Delta^{i}-1}\right]\exp{\left(-\sum_{j}\sum_{i<j}\alpha^{i}\alpha^{j}(x^{ij})^{2}\right)}                             
 \label{T1Veqn10}    
\end{eqnarray}
To proceed further, we now use the inverse Mellin transform representation of the exponential function 
\begin{equation}
e^{-x}=\frac{1}{2\pi i}\int_{c-i\infty}^{c+i\infty}ds \ \Gamma(s)x^{-s}           
\label{T1Veqn12} 
\end{equation}
\mvnote{Here $c$ is a real number greater than or equal to zero (if $c=0$, then  the contour of integration has a dent at the origin so as to put the pole at the origin on the left). The contour can be shifted to the right freely as all the poles are on negative real axis ( note that the poles of $\Gamma(s)$ are at $0$ and all negative integers). This freedom in shifting the contour (or equivalently, the freedom in the choice of $c$) turns out to be very crucial as we shall see below.}

\vspace*{.07in}\mvnote{Using ~\eqref{T1Veqn12} for the exponential factor in ~\eqref{T1Veqn10}, we obtain,}
\begin{eqnarray}
I=\prod_{j}\prod_{i<j}\left[\int_{c^{ij}-i\infty}^{c^{ij}+i\infty} [ds^{ij}](x^{ij})^{-2s^{ij}}\Gamma(s^{ij})\right] \prod_{i=1}^{N}\left[\int_{0}^{\infty}d\alpha^{i}(\alpha^{i})^{\rho^{i}-1}\right]
 \label{T1Veqn13}
\end{eqnarray}
\mvnote{where, the $s^{ij}$ (corresponding to $x^{ij}$) are our Mellin variables and $c^{ij}$ are real numbers greater than zero. Also,}
\begin{equation}
\rho^{i}\equiv\Delta^{i}-\sum_{\substack{j={1}\\j\not=i}}^{N}s^{ij}\quad,\qquad\qquad [ds^{ij}]\equiv \frac{ds^{ij}}{2\pi i}
 \label{T1Veqn14}
\end{equation}
\mvnote{If $\rho^i$ had any real part, the integration over the Schwinger parameters $\alpha^i$ in \eqref{T1Veqn13} will give divergent result. However, as explained in the appendix \ref{mellin_delta}, the integrals $\int_{0}^{\infty}d\alpha^{i}(\alpha^{i})^{\rho^{i}-1}$ behave as delta function inside the contour integration provided the real part of the exponent $\rho^i$ is zero along the contour. More explicitly, as shown in the appendix \ref{mellin_delta}, we have the following result}
 \be
  \int_{c-i\infty}^{c+i\infty}[ds]f(s)\int_0^\infty dt\ t^{s_0-s-1}=\int_{c-i\infty}^{c+i\infty}[ds]f(s)\left(2\pi i\delta(s-s_o)\right)
  \label{addmede}
  \ee
\mvnote{if we choose $c=\mbox{Re}(s_0)$ (so that the real part of the exponent $s_0-s$ is zero along the contour). }

\vspace*{.07in}\mvnote{Thus, for the Schwinger parameter integrals in \eqref{T1Veqn13} to be well defined, we need to ensure that the real part of the exponents $\rho^i$ vanish along the contour of integration. Using the expression of $\rho^i$ given in \eqref{T1Veqn14}, this means that we need to choose the set $\{c_{ij}\}$ in such a way that they satisfy}
\mvnote{\begin{eqnarray}
\sum_{\substack{j=1\\ j\neq i}}^{N}c^{ij}=\Delta^{i}          \label{addedlater3}
\end{eqnarray}}
\mvnote{We observe from \eqref{T1Veqn13} that \eqref{addedlater3} is the same set of constraints that $s^{ij}$ must satisfy if the expression in \eqref{T1Veqn13} has to transform correctly under the conformal transformations. Therefore we can infer that a solution to \eqref{addedlater3} exists (or else it would lead to a contradiction) and the Schwinger parameter integrals in \eqref{T1Veqn13} are well defined.}

\vspace*{.07in}\mvnote{Using \eqref{addmede}, the expression \eqref{T1Veqn13} becomes}
\begin{eqnarray}
I=\prod_{j}\prod_{i<j}\left[ \int_{c^{ij}-i\infty}^{c^{ij}+i\infty}[ds^{ij}](x^{ij})^{-2s^{ij}}\Gamma(s^{ij})\right]\left[\prod_{i}2\pi i\delta\left(\Delta^{i}-\sum_{\substack{j\not=i}}s^{ij}\right)\right]
\label{T1Veqn16}
\end{eqnarray}
As discussed above, the constraints $\rho_i=0$ (enforced by the delta functions) are precisely the constraints on the Mellin variables discussed in section \ref{sec:mellin}. These constraints originate from the fact that the position space correlation function is covariant under conformal transformations. We also note that these constraints reduce the number of independent Mellin variables from $N(N-1)/2$ to $N(N-3)/2$ which is also the number of independent cross-ratios for $N$ points\footnote{This is true in generic dimensions. In special cases, the number of independent cross ratios may be different}. 

\vspace*{.06in}We can now read off the Mellin ampltitude corresponding to the Feynman diagram in Figure \ref{T1Vdiag1}. Comparing \eqref{T1Veqn16} with the defining expression of Mellin amplitude \eqref{Mellin1}, \mvnote{we find that the Mellin amplitude for contact interaction is just 1.}

\vspace*{.06in}A careful look at \eqref{T1Veqn16} tells us that the $N$ delta functions force the $(x^{ij})^{-2s^{ij}}$ terms to combine and form $\frac{N(N-3)}{2}$ cross ratios between the external vertices $x^{i}$ and some extra factors that give appropriate transformation properties to the position space correlator.


\subsection{Tree With One Internal Propagator}                           
\label{T2}

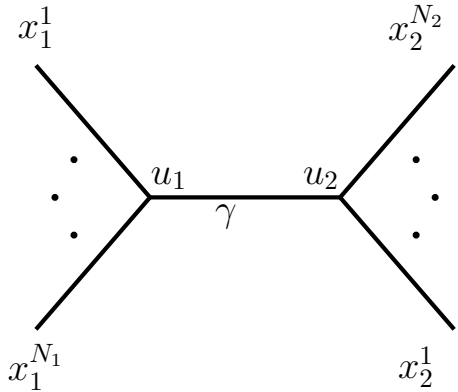
\begin{figure}[h]
\begin{center}
\begin{tikzpicture}[scale=.5]
\draw [-] [ultra thick](-3,-3.5) -- (0,0);
\draw [-] [ultra thick](-3,3.5) -- (0,0);
\draw [-] [ultra thick](5,0) -- (0,0);
\draw [-] [ultra thick](8,-3.5) -- (5,0);
\draw [-] [ultra thick](8,3.5) -- (5,0);
\draw (.5,.5) node{\Large $u_1$};
\draw (4.5,.5) node{\Large $u_2$};
\draw (-3,-4.5) node{\Large $x_1^{N_1}$};
\draw (7,-4.5) node{\Large $x^{1}_2$};
\draw (7,4.5) node{\Large $x_2^{N_2}$};
\draw (-3,4.5) node{\Large $x_1^1$};
\draw (2,-.5) node{\Large $\gamma$};
\begin{scope}[shift={(-3,0)}] 
\filldraw [ultra thick] (1,1) circle (1pt);
\filldraw [ultra thick] (.5,0) circle (1pt);
\filldraw [ultra thick] (1,-1) circle (1pt);
\end{scope}
\begin{scope}[shift={(6,0)}] 
\filldraw [ultra thick] (1,1) circle (1pt);
\filldraw [ultra thick] (1.5,0) circle (1pt);
\filldraw [ultra thick] (1,-1) circle (1pt);
\end{scope}
\end{tikzpicture}
\end{center}
\caption{Two vertex}
\label{diag:twov1}
\end{figure}

The next Feynman diagram that we consider (Figure \ref{diag:twov1}) involves two internal vertices connected by an internal propagator. This example will give us the expression for the scalar propagator in Mellin space. $\gamma$ denotes the scaling dimension of the internal propagator.
\\
The position space expression for this diagram is given by
\begin{eqnarray}
I=\int \mathcal{D}u_1 \ \mathcal{D}u_2\Biggl[\prod_{i\in1}(x_1^i-u_1)^
{-2\exc_1^i}\Gamma(\exc_1^i)
\prod_{j\in2}(x_2^j-u_2)^{-2\exc_2^j}\Gamma(\exc_2^j)
(u_2-u_1)^{-2\gamma}\Biggl]
\non
\label{twov1}
\end{eqnarray} 
\vspace*{.06in}The conformality conditions for the two interaction vertices in this diagram are,
\begin{eqnarray}
\sum\limits_{i\in1}\Delta_1^i+\inc=D\quad,\qquad \qquad \sum\limits_{i\in2}\Delta_2^i+\inc=D
\label{confor_2}
\end{eqnarray}
\vspace*{.06in}We again use the identity \eqref{T1Veqn2} and introduce the Schwinger parameters for each propagator (internal as well as external)
\begin{eqnarray}
I&=&
\Biggl[\prod\limits_{i\in1}\int_0^\infty d\alpha_1^i(\alpha_1^i)
^{\exc_1^i-1}
\prod\limits_{j\in2}\int_0^\infty d\alpha_2^j
(\alpha_2^j)^{\exc_2^j-1}
\frac{1}{\Gamma(\gamma)}\int_0^\infty dt\;t^{\inc-1}\Biggl]\int \mathcal{D}u_1 \ \mathcal{D}u_2
\nonumber\\
&&\hspace*{.4in}\exp\Biggl(-\sum\limits_{i\in1}\alpha_1^i(x_1^i-u_1)^{2}
-\sum_{j\in2}\alpha_2^j(x_2^j-u_2)^{2}-t(u_2-u_1)^2\Biggl)
\non
\label{twov2}
\end{eqnarray}
where $t$ is the Schwinger parameter for the internal propagator. 

\vspace*{.06in}Performing the $u_1$ integration, we obtain
\begin{eqnarray}
 I=&&\frac{1}{2\Gamma(\inc)}
\prod\limits_{a=1}^2\prod\limits_{i\in a}
\left(\int 
d\alpha_a^i(\alpha_a^i)^{\exc_a^i-1}
\right)
\int_0^\infty dt\;t^{\inc-1}
\int \mathcal{D}u_2\
\exp\Biggl(-\sum_{j\in2}(x_2^j-u_2)^{2}\Biggl)
\nonumber\\
&&\Bigl(\sum\limits_{i\in1}\alpha_1^i+t\Bigl)^{-D/2}\exp\Biggl(-
\Bigl(\sum\limits_{i\in1}\alpha_1^i+t\Bigl)^{-1}
\biggl\{\sum_{(i,j)\in 1}\alpha_1^i\alpha_1^j (x_{11}^{ij})^2+t\sum_{i\in1}
\alpha_1^i(x_{1}^i-u_2)^2\biggl\}\Biggl)
\non\\
\label{twov3}
\end{eqnarray}
 Next, we insert the partition of unity 
\begin{eqnarray}
1=\int_0^\infty dy\:\delta\Bigl(y-\sum_{i\in1}\alpha_1^i-t\Bigl)\nonumber
\end{eqnarray}
 in the integral of \eqref{twov3}, rescale the Schwinger parameters 
\begin{eqnarray}
\alpha_1^i\rightarrow \sqrt{y}\ \alpha_1^i\qquad,\qquad t\rightarrow \sqrt{y} \ t
\label{twov4}
\end{eqnarray}
and perform the integration over the variable $y$ using the delta function. The result is 
\begin{eqnarray}
I&=&\frac{1}{\Gamma(\inc)}
\prod\limits_{a=1}^2\prod\limits_{i\in a}\Biggl[\int_0^\infty d\alpha_a^i(\alpha_a^i)^{\exc_a^i-1}\Biggl]
\int_0^\infty dt\;t^{\inc-1}\Bigl(\sum_{i\in1}\alpha_1^i+t\Bigl)^{\sum\limits_{i\in1}\Delta_1^i+\inc-D}
\nonumber\\
&&\exp\Biggl(-\sum_{(i,j)\in 1}\alpha_1^i\alpha_1^j(x_{11}^{ij})^{2}\Biggl)
\int\mathcal{D}u_2\exp\Biggl(-\sum_{j\in2}\alpha_2^j(x_2^j-u_2)^{2}
-t\sum_{i\in1}\alpha_1^i(x_{i}-u_2)^2\Biggl)
\nonumber
\end{eqnarray}
\vspace*{.06in}Next, we perform the $u_2$ integration, insert the following partition of unity in the integral
 \begin{eqnarray}
1=\int_0^\infty dy\:\delta\Bigl(y-\sum_{i\in2}\alpha_2^i-t\sum_{i\in1}\alpha_1^i\Bigl)\ \ ,
\nonumber
\end{eqnarray}
carry out similar rescalings as in \eqref{twov4} (but this time, with the variables $\alpha_2^i$ and $t$) and perform the integration over the auxiliary variable $y$. This gives,
\begin{eqnarray}
I&=&\frac{1}{\Gamma(\inc)}
\prod\limits_{i\in1}\left[\int_0^\infty d\alpha_1^i(\alpha_1^i)^{\exc_1^i-1}\right]
\prod\limits_{j\in2}\left[\int_0^\infty d\alpha_2^j(\alpha_2^j)^{\exc_2^j-1}\right]
\int_0^\infty dt\;t^{\inc-1}
\nonumber\\
&&\exp\Biggl(-(1+t^2)\sum_{(i,j)\in 1}\alpha_1^i\alpha_1^j(x_{11}^{ij})^{2}
-\sum_{(i,j)\in2}
\alpha_2^i\alpha_2^j(x_{22}^{ij})^2-t\sum_{i\in1}\sum_{j\in2}\alpha_1^i\alpha_2^j(x_{12}^{ij})^2\Biggl)
\non\\
&&\left(\Bigl(1+t^2\Bigl)\sum_{i\in1}\alpha_1^i+t\sum_{i\in2}
\alpha_2^i\right)^{\sum\limits_{i\in 1}\Delta_1^i+\inc-D}
\Bigl(\sum_{i\in2}\alpha_2^i+t\sum_{i\in1}\alpha_1^i\Bigl)
^{\sum\limits_{i\in2}\Delta_2^i+\inc-D}\non
\label{twov5}
\end{eqnarray}
We impose the conformality conditions \eqref{confor_2}, and then use the identity \eqref{T1Veqn12} for each exponential factor. After some rearrangement, we obtain, 
\begin{eqnarray}
I&=&\frac{1}{\Gamma(\inc)}
\prod\limits_{(i,j)\in 1+2}
\Biggl(
\int_{c_{ab}^{ij}-\iifty}^{c_{ab}^{ij}+\iifty}[{d\mv_{ab}^{ij}}]\Gamma(\mv_{ab}^{ij})
\Bigl((x_{ab}^{ij})^{2}\Bigl)^{-\mv_{ab}^{ij}}
\Biggl)\prod\limits_{i\in1}\Biggl(\int_0^\infty d\alpha_1^i(\alpha_1^i)^{\rho_1^i-1}\Biggl)
\nonumber\\
&&
\hspace*{1in}\prod\limits_{j\in2}\Biggl(\int_0^\infty d\alpha_2^j(\alpha_2^j)^{\rho_2^j-1}\Biggl)\int_0^\infty dt\;t^{\inc-s_{12}-1}(1+t^2)^{-s_{11}}
\label{twov8}
\end{eqnarray}
where,
\begin{eqnarray*}
\rho_1^{i}&\equiv&\Delta_1^i-\sum\limits_{j\in1}\mv_{11}^{ij}-\sum\limits_{j\in2}\mv_{12}^{ij}\quad,\qquad\quad (i\in 1)
\\
\rho_2^{i}&\equiv&\Delta_2^i-\sum\limits_{j\in2}\mv_{22}^{ij}-\sum\limits_{j\in1}\mv_{12}^{ji}\quad,\qquad\quad (i\in 2)
\\
s_{12}&\equiv&\sum\limits_{i\in1}\sum\limits_{j\in2}\mv_{12}^{ij}\quad;\qquad s_{aa}\equiv\sum\limits_{1\le i<j\le N_a}\mv_{aa}^{ij}\ \ ,\quad a=1,2
\end{eqnarray*}
Once again, we choose $c_{ab}^{ij}$ appropriately so that the integrals over the Schwinger parameters $\alpha_1^i$ and $\alpha_2^i$ act as delta functions in the Mellin space (as explained in the previous section). This means that the Mellin variables satisfy the constraints
\begin{eqnarray}
\rho_1^i=0=\rho_2^i \;\;\;\;\;\;\;\;\;\;\;\;\;\;\; \forall \ i
\label{twov6}
\end{eqnarray}
These reduce the number of independent Mellin variables from $(N_1+N_2)(N_1+N_2-1)/2$ to $(N_1+N_2)(N_1+N_2-3)/2$. Summing over $i$ and using the conformality conditions \eqref{confor_2} gives useful relations between the Mellin variables
\begin{eqnarray}
\sum\limits_{i\in1}\Delta_1^i=2s_{11}+s_{12}=D-\inc\quad,\qquad \sum\limits_{i\in2}\Delta_2^i=2s_{22}+s_{12}=D-\inc
\label{twov9}
\end{eqnarray} 
By comparing \eqref{twov8} with the definition of Mellin amplitude \eqref{Mellin1} (taking into account the constraints \eqref{twov6}), we can easily read off the Mellin amplitude to be

\begin{eqnarray}
M(\mv_{12})=\frac{1}{\Gamma(\inc)}\int_0^\infty dt\;t^{\inc-s_{12}-1}(1+t^2)^{-s_{11}}=\frac{1}{2\Gamma(\inc)}\beta\Bigl(\frac{\inc-s_{12}}{2},\frac{D}{2}-\inc\Bigl )                \label{twov10}
\end{eqnarray}
where we have used \eqref{twov9} to simplify the arguments of beta function.

\vspace*{.06in}The physical interpretation of the amplitude is clear. We can identify the beta function to be the Mellin space propagator. Moreover, the poles of the beta function have clear physical interpretation. At this stage, it is convenient to introduce dual Mellin momenta. If the Mellin momentum flowing into the internal propagator through the external vertex $x_a^i$ be $k_a^i$ (where we have suppressed the dual spacetime index), then the full momentum propagating through the internal propagator is 
\begin{eqnarray}
k=\sum_{i \in 1}k^i_1=
-\sum_{i \in 2 }k^i_2
\non
\end{eqnarray}
and the kinematical variable entering into the propagator is 
\begin{eqnarray}
\mv_{12}=\sum_{i\in 1}\sum_{j \in 2}\mv_{12}^{ij}=\left(\sum_{i\in 1}k^i_1\right)\cdot  \left(\sum_{j \in 2}k_2^j\right)=-k^2
\non
\end{eqnarray}
This means that the poles of the propagator appear at particular values of the $k^2$, namely 
\be
s_{12}=-k^2=\gamma+2n\qquad;\quad n=0,1,2,\cdots
\label{poles11}
\ee
\mvnote{These poles correspond to the primary and its leading twist descendants ($n=0$) and the satellite ($n>0$) propagating states\footnote{\mvnote{As shown by Mack \cite{Mack1}, the poles of the Mellin amplitude occur at $s=\gamma-\ell +2n\;\;,\; (n=0,1,2\cdots )$. The first pole $(n=0)$ corresponds to the exchanged primary operator and all its leading twist descendants (i.e. those operators in the conformal multiplet whose dimensions and spins keep $\gamma-\ell$ fixed). The higher poles ($n>0$) correspond to the satellite poles. The above result \eqref{poles11} is consistent with this expectation.}}. }


\subsection{Tree With Two Internal Propagators}
\label{sec:3vertree}
In order to check our interpretation of the result \eqref{twov10} as the propagator in Mellin space, we consider one more example before generalising to arbitrary tree level diagrams. We consider a Feynman diagram with two internal propagators \mvnote{(see figure \ref{diag:t3}).
\begin{figure}[h]
\begin{center}
\begin{tikzpicture}[scale=.5]
\draw [-] [ultra thick](-3,-3.5) -- (0,0);
\draw [-] [ultra thick](-3,3.5) -- (0,0);
\draw [-] [ultra thick](10,0) -- (0,0);
\draw [-] [ultra thick](13,-3.5) -- (10,0);
\draw [-] [ultra thick](13,3.5) -- (10,0);
\draw [-] [ultra thick](2,-3.5) -- (5,0);
\draw [-] [ultra thick](8,-3.5) -- (5,0);
\draw (.5,.5) node{\Large $u_1$};
\draw (4.5,.5) node{\Large $u_2$};
\draw (9.5,.5) node{\Large $u_3$};
\draw (-3,-4.5) node{\Large $x_1^{N_1}$};
\draw (13,-4.5) node{\Large $x_3^{N_3}$};
\draw (13,4.5) node{\Large $x^{1}_3$};
\draw (2,-4.5) node{\Large $x^{1}_2$};
\draw (8,-4.5) node{\Large $x_2^{N_2}$};
\draw (-3,4.5) node{\Large $x_1^1$};
\draw (2,-.5) node{\Large $\gamma_{12}$};
\draw (7,-.5) node{\Large $\gamma_{23}$};
\begin{scope}[shift={(-3,0)}] 
\filldraw [ultra thick] (1,1) circle (1pt);
\filldraw [ultra thick] (.5,0) circle (1pt);
\filldraw [ultra thick] (1,-1) circle (1pt);
\end{scope}
\begin{scope}[shift={(5,0)}] 
\filldraw [ultra thick] (-1,-2) circle (1pt);
\filldraw [ultra thick] (0,-2.5) circle (1pt);
\filldraw [ultra thick] (1,-2) circle (1pt);
\end{scope}
\begin{scope}[shift={(11,0)}] 
\filldraw [ultra thick] (1,1) circle (1pt);
\filldraw [ultra thick] (1.5,0) circle (1pt);
\filldraw [ultra thick] (1,-1) circle (1pt);
\end{scope}
\end{tikzpicture}
\end{center}
\caption{Three vertex tree}
\label{diag:t3}
\end{figure}
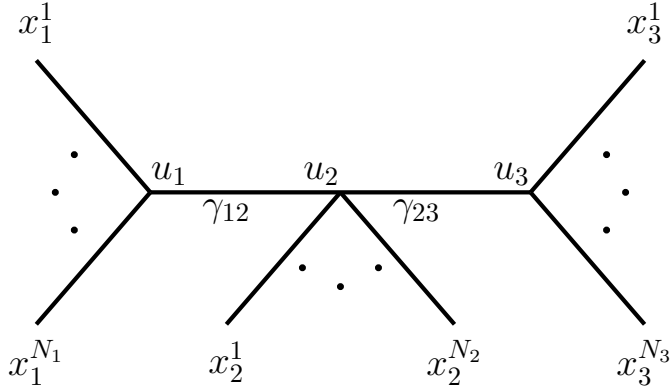}\\
The position space expression for this is given by
\begin{eqnarray}
I&=&\int \mathcal{D}u_1\mathcal{D}u_2\mathcal{D}u_3
\Biggl[
\prod\limits_{i\in 1}\left\{(x^i_1-u_1)^{-2\exc^i_1}\Gamma(\exc^i_1)\right\}
\prod\limits_{i\in 2}\left\{(x^i_2-u_2)^{-2\exc^i_2}\Gamma(\exc^i_2)\right\}
\nonumber\\
&&
\hspace*{.3in}\times\ \prod\limits_{i\in 3}\left\{(x^i_3-u_3)^{-2\exc^i_3}\Gamma(\exc^i_3)\right\}(u_2-u_1)^{-2\inc_{12}}
(u_2-u_3)^{-2\inc_{23}}
\Biggl]
\label{illus1}
\end{eqnarray}
In this case, the conformality conditions are 
\begin{eqnarray}
\sum\limits_{i\in a}\exc^i_a&=&D-\inc_{a,a+1}-\inc_{a-1,a}\quad,\qquad 1\le a\le 3
\non\label{threev8}
\end{eqnarray}
where, $\gamma_{01}=0=\gamma_{34}$.

\vspace*{.07in}For extracting the Mellin amplitude, we follow the same strategy as in the previous examples. However, now we have to choose an ordering of the vertices $u_a$ for conducting the manipulations. All the choices lead to the same result eventually\footnote{The different choices for this ordering lead to integrals over the Schwinger parameters which are not manifestly equal. For diagrams with higher number of interaction vertices, it can often be difficult to show that these different integrals corresponding to the same Feynman diagram are all equal.}. We follow the order $u_1\rightarrow u_2\rightarrow u_3$. The final result turns out to be
\begin{eqnarray}
I&=&\prod\limits_{a=1}^3\prod\limits_{b=a}^3\Biggl(
\prod\limits_{ (i, j)\in a+b}\int_{c_{ab}^{ij}-\iifty}^{c_{ab}^{ij}+\iifty}
\left[d\mv_{ab}^{ij}\right]
\Gamma(\mv_{ab}^{ij})\bigl(x_{ab}^{ij}\bigl)^{-2\mv_{ab}^{ij}}\Biggl)
\prod\limits_{a=1}^{3}\Biggl(\prod\limits_{i\in a}\int_0^\infty d\alpha_a^i(\alpha_a^i)^{\rho_a^i-1}\Biggl)
M(\mv_{ab})
\nonumber
\end{eqnarray}
where,
\begin{eqnarray}
\rho_a^{i}&\equiv&\exc_a^i-\sum\limits_{j\in a}\mv_{aa}^{ij}-\sum_{\substack{b={1}\\b\not=a}}^{3}\biggl(\sum\limits_{j\in b}
\mv_{ab}^{ij}\biggl)\quad,\qquad 1\le a\le 3
\non\label{treev4}
\end{eqnarray}
Again, the integration over the variables $\alpha_a^i$ impose the constraints $\rho_a^i=0$ which can be re-written as (using the conformality conditions)
\begin{eqnarray}
\sum\limits_{i\in a}\exc^i_a&=&
2\mv_{aa}+\sum_{\substack{b={1}\\b\not=a}}^{3}\mv_{ab}=D-\inc_{a-1,a}-\inc_{a,a+1}\quad,\qquad 1\le a\le 3
\non
\label{threev5}
\end{eqnarray}
Due to these constraints, the number of independent Mellin variables are only $N(N-3)/2$ (where $N$ is total number of external states).
\vspace*{.06in}The Mellin amplitude is given by
\begin{eqnarray}
M(s_{ab})&=&\frac{1}{\Gamma(\inc_{12})\Gamma(\inc_{23})}
\int_0^\infty dt_{12}(t_{12})^{\inc_{12}-\mv_{12}-\mv_{13}-1}
\int_0^\infty dt_{23}(t_{23})^{\inc_{23}-\mv_{13}-\mv_{23}-1}
\nonumber\\&&
\bigl(1+t_{12}^2(1+t_{23}^2)\bigl)^{-\mv_{11}}
\Bigl(1+t^2_{23}\Bigl)^{-\mv_{22}-\mv_{12}}
\label{threev7}\\
&=&\Biggl[\frac{1}{2\Gamma(\inc_{12})}\beta\Bigl(\frac{\inc_{12}-s_{12}-s_{13}}{2},\frac{D}{2}-\inc_{12}\Bigl)\Biggl]\Biggl[\frac{1}{2\Gamma(\inc_{23})}\beta\Bigl(\frac{\inc_{23}-s_{23}-s_{13}}{2},\frac{D}{2}-\inc_{23}\Bigl)\Biggl]
\nonumber
\end{eqnarray}

\vspace*{.06in}This result, being a product of two beta functions with appropriate arguments, is consistent with our interpretation of the Mellin space propagator \eqref{twov10}. Moreover, the poles of the propagator occur when the negative of the total Mellin momenta squared flowing through it is equal to the conformal dimension of the primary and descendants. This is easily seen by introducing the dual Mellin momenta and writing the arguments of beta functions in terms of these momenta.


\section{General Tree Level Feynman Diagrams}
\label{sec:diag}
We now consider general tree level Feynman diagrams. We shall show that the Mellin amplitude for an arbitrary tree diagram is given by the product over internal propagators. For each internal propagator, we obtain a factor of beta function with appropriate arguments consistent with the examples considered in the previous section. 

\vspace*{.07in}In subsection \ref{sec: diag_rules}, we present a diagrammatic algorithm to write down the Mellin amplitude (in terms of integrals over the Schwinger parameters) for any diagram involving only scalar operators. In subsection \ref{sec:simtree}, we consider a tree diagram with $n$ internal vertices such that one can go from one end of the diagram to the other end without encountering any branches (figure \ref{ntree}). Finally, in subsection \ref{TG}, we consider a completely general tree diagram. 

\subsection{Diagrammatic Rules for Writing Mellin Amplitude}
\label{sec: diag_rules}

In this subsection, we present a diagrammatic technique which will be helpful in directly writing down the Mellin space amplitudes as integrals over the Schwinger parameters. These rules can be used for any tree as well as loop diagrams and will allow us to avoid going through all the algebraic manipulations, as described in the examples of the section \ref{TL}. Although these rules are quite crucial in our derivation of the Mellin amplitude for a general tree level Feynman diagram, a casual reader may skip this subsection and commence reading from subsection \ref{sec:simtree}. 

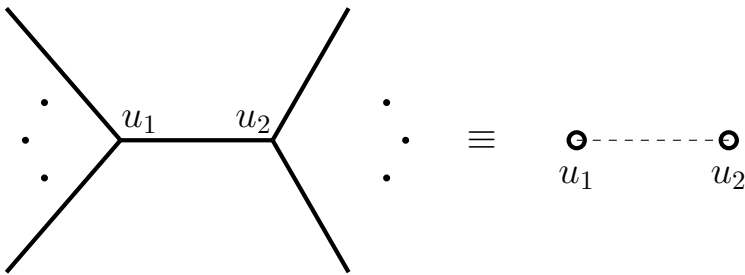
\begin{figure}[h]
\begin{center}
\begin{tikzpicture}[scale=.5]
\draw [-] [ultra thick](-3,-3.5) -- (0,0);
\draw [-] [ultra thick](-3,3.5) -- (0,0);
\draw [-] [ultra thick](4,0) -- (0,0);
\draw [-] [ultra thick](6,-3.5) -- (4,0);
\draw [-] [ultra thick](6,3.5) -- (4,0);
\draw (.5,.5) node{\Large $u_1$};
\draw (3.5,.5) node{\Large $u_2$};
\draw (9.5,0) node{\Large $\equiv$};
\draw [dashed] (12,0) -- (16,0);
\draw [ultra thick] (12,0) circle (6pt);
\draw [ultra thick] (16,0) circle (6pt);
\draw (12,-1) node {\Large $u_1$};
\draw (16,-1) node {\Large $u_2$};
\begin{scope}[shift={(-3,0)}] 
\filldraw [ultra thick] (1,1) circle (1pt);
\filldraw [ultra thick] (.5,0) circle (1pt);
\filldraw [ultra thick] (1,-1) circle (1pt);
\end{scope}
\begin{scope}[shift={(6,0)}] 
\filldraw [ultra thick] (1,1) circle (1pt);
\filldraw [ultra thick] (1.5,0) circle (1pt);
\filldraw [ultra thick] (1,-1) circle (1pt);
\end{scope}

\end{tikzpicture}
\end{center}
\caption{Skeleton of the single propagator diagram}
\label{diag:step1}
\end{figure}

\vspace*{.07in} For developing these rules, we shall use a simplified way to represent the Feynman diagrams. In our diagrammatic algorithm, the external lines in a Feynman diagram would not be playing any significant role.  Hence, to simplify the diagrammatic representation, we represent the set of external lines attached to an interaction vertex by a small hollow circle at the vertex and the internal propagator by dashed lines. We call this the {\it skeleton of the Feynman diagram}. The skeleton for the single propagator Feynman diagram we considered earlier, is shown in Figure \ref{diag:step1}. Note that this way of representing a Feynman diagram is insensitive to the number of external legs attached to any given interaction vertex. 


\subsubsection{\mvnote{Illustrating the rules}}

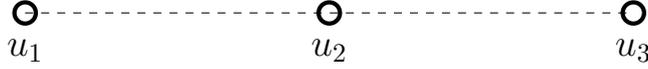
\begin{figure}[h]
\begin{center}
\begin{tikzpicture}[scale=1]
\draw [dashed]  (-4,0) -- (4,0);
\draw [ultra thick] (-4,0) circle (4pt);
\draw [ultra thick] (0,0) circle (4pt);
\draw [ultra thick] (4,0) circle (4pt);
\draw (-4,-.5) node {\Large $u_1$};
\draw (0,-.5) node {\Large $u_2$};
\draw (4,-.5) node {\Large $u_3$};
\end{tikzpicture}
\end{center}
\caption{Tree level three vertex }
\label{diag:3tree}
\end{figure}

\mvnote{In this section, we shall consider an explicit example to understand the diagrammatic rules. We consider the two propagator case of section \ref{TL} for this purpose. The skeleton of this Feynman diagram is shown in Figure \ref{diag:3tree}. Considering this example serves a two-fold purpose. Apart from being an explicit (and simple) example of the application of the rules, it also helps us understand the origin of the rules. The essential idea is to represent the steps of derivation leading to the Mellin amplitude as a series of diagram. The components of the diagram are assigned some weight factors. The Mellin amplitude can be written in terms of the weight factors of the final diagram obtained after integrating over all the position space vertices.}

\vspace*{.06in}\mvnote{We start with the position space expression for this diagram which is given in equation \eqref{illus1}. For integrating over the interaction vertices, we choose the ordering $u_1\rightarrow u_2\rightarrow u_3$. 
We first consider the effect of integration over the $u_1$ variable. When $u_{1}$ is integrated over, we get terms of the form $\alpha_{1}^{i}\alpha_{1}^{j}(x_{1}^{i}-x_{1}^{j})^{2}$ and $\alpha_{1}^{i}t_{12}(x_{1}^{i}-u_{2})^{2}$ in the exponent. The factors of $\alpha^{i}$ eventually do not contribute to the Mellin amplitude (their role is in providing the delta function constraints on the Mellin variables as seen in examples of previous section). After using the Cahen Mellin identity for the exponentials, the coefficient of $\alpha_1^i\alpha_1^j(x_{1}^{i}-x_{1}^{j})^2$ essentially becomes the part of Mellin amplitude. Keeping this in mind, we assign a weight $1$ to the factor $(x_{1}^{i}-x_{1}^{j})^2$. We also assign a weight of $t_{12}$ with the factor $(x_{1}^{i}-u_{2})^2$ (the reason for this will become clear shortly). }

\begin{figure}[h]
\begin{center}
\begin{tikzpicture}[scale=.8,transform shape]
\draw[thick, black] (0,0) -- (3.9,0);
\draw[thick, black,dashed] (4.1,0) -- (8,0);
\draw [ultra thick] (8,0) circle (5pt);
\draw [ultra thick] (4,0) circle (5pt);
\filldraw[fill=white, draw=black]  (-.3,.28,0.5) circle (0.5cm);
\node at (-1.4,0) {\Large$1$};
\draw (2,.6) node {\Large $t_{12}$};
\end{tikzpicture} 
\end{center}
\caption{Diagrammatic representation of integration over vertex 1}
\label{diag: int_1}
\end{figure}
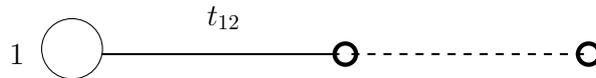

\vspace*{.06in}The statements made in the previous paragraph can be nicely captured by a diagrammatic means. We take the diagram in figure \ref{diag:3tree} and replace the small hollow circle associated to the vertex $u_1$ with a bigger circle and the dashed lines connecting the adjacent un-integrated vertex with solid lines. We associate a weight $1$ with this bigger circle and a weight $t_{12}$ with the solid line (which is the Schwinger parameter associated with the line joining the vertices 1 and 2). This has been shown in Figure \ref{diag: int_1} (from now on, we won't write the vertex indices $u_a$).

\begin{figure}[h]
\begin{center}
\begin{tikzpicture}[scale=.8,transform shape]
\draw[thick, black] (0,0) -- (3.9,0);
\draw[thick, black] (4.1,0) -- (8,0);
\draw [ultra thick] (8.1,0) circle (5pt);
\draw[thick,black]  (-.5,-.1) circle (0.5cm);
\draw[thick,black]  (-1,-.1) circle (1cm);
\filldraw[fill=white, draw=black]  (4.2,.65,0.5) circle (0.5cm);
\draw [ thick,black] (8,0) to[out=250,in=290]  (0,0);
\node at (-1.3,0) {\Large$1$};
\node at (4,1.3) {\Large$1$};
\node at (-2.5,0) {\Large$t_{12}^2$};
\draw (2,.6) node {\Large $t_{12}$};
\draw (6,.6) node {\Large $t_{23}$};
\node at (4,-2.5) {\Large$t_{12}t_{23}$};
\end{tikzpicture} 
\end{center}
\caption{Diagrammatic representation of integration over second vertex}
\label{diag: int_2}
\end{figure}
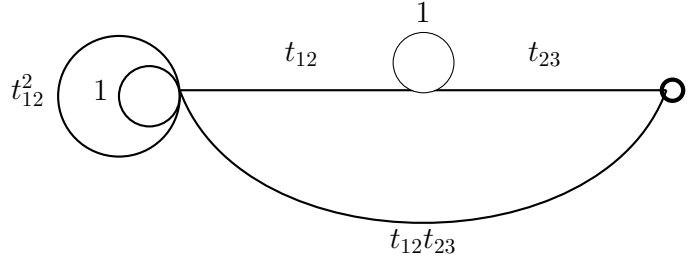

\vspace*{.06in}\mvnote{Next, we perform the integration over the $u_2$ vertex. This gives rise to terms of the form $(1+t_{12}^{2})\alpha_{1}^{i}\alpha_{1}^{j}(x_{1}^{i}-x_{1}^{j})^{2}$, $t_{12}\alpha_{1}^{i}\alpha_{2}^{j}(x_{1}^{i}-x_{2}^{j})^{2}$, $\alpha_{2}^{i}\alpha_{2}^{j}(x_{2}^{i}-x_{2}^{j})^{2}$, $\alpha_{2}^{i}t_{23}(x_{2}^{i}-u_{3})^{2}$ and $\alpha_{1}^{i}t_{12}t_{23}(x_{1}^{i}-u_{3})^{2}$ in the exponent. Keeping in mind that the coefficients of $\alpha_a^i\alpha_b^j(x_{a}^{i}-x_{b}^{j})^{2}$ eventually become part of the Mellin amplitude, we diagrammatically represent this step by replacing the small circle around the $u_2$ vertex with a bigger circle, replace the dashed line connecting it with $u_3$ vertex by a solid line and making one more circle around the $u_1$ vertex. We also connect the vertices $u_1$ and $u_3$ by a different solid line. From the terms just mentioned, we see that we need to associate a weight $1$ with the circle around $u_2$ vertex, a weight $t_{23}$ with the solid line connecting $u_2$ and $u_3$ vertices, a weight $t_{12}t_{23}$ with the solid line connecting $u_1$ and $u_3$ and a weight $t_{12}^2$ with the new circle around $u_1$. This step, combined with the first step, can be represented as in Figure \ref{diag: int_2}.}

\mvnote{At this stage, we can state our strategy as follows: as mentioned above, the coefficients of $\alpha_a^i\alpha_b^j(x_{a}^{i}-x_{b}^{j})^{2}$ in exponent in the final expression becomes the part of the Mellin amplitude. We have chosen to represent the coefficients for the case $a=b$ (i.e. $\alpha_a^i\alpha_a^j(x_{a}^{i}-x_{a}^{j})^{2}$ ) by associating the weight factors to the circles drawn around the vertex $u_a$. On the other hand, the coefficients for the case $a\not=b$ (i.e. $\alpha_a^i\alpha_b^j(x_{a}^{i}-x_{b}^{j})^{2}$ ) are represented by associating the weight factors to the solid line connecting the vertices $u_a$ and $u_b$. }

\begin{figure}[h]
\begin{center}
\begin{tikzpicture}[scale=.8,transform shape]
\draw[thick, black] (0,0) -- (3.9,0);
\draw[thick, black] (4.1,0) -- (8,0);
\draw[thick,black]  (-.5,-.1) circle (0.5cm);
\draw[thick,black]  (-1,-.1) circle (1cm);
\draw[thick,black]  (-2,-.1) circle (2cm);
\draw[thick,black]  (8.5,0) circle (.5cm);
\draw[thick,black]  (4.1,.5) circle (0.5cm);
\draw[thick,black]  (4.1,1) circle (1cm);
\draw [ thick,black] (8,0) to[out=250,in=290]  (0,0);
\draw [ thick,black] (0,0) to[out=30,in=160]  (4,0);
\node at (-1.3,0) {\Large$1$};
\node at (-1.3,0) {\Large$1$};
\node at (4,1.3) {\Large$1$};
\node at (9.3,0) {\Large$1$};
\node at (4,2.6) {\Large$t_{23}^2$};
\node at (-5,0) {\Large$t_{12}^2t_{23}^2$};
\node at (-2.5,0) {\Large$t_{12}^2$};
\draw (2,-.6) node {\Large $t_{12}$};
\draw (6,-.6) node {\Large $t_{23}$};
\node at (1.6,1.1) {\Large$t_{12}t_{23}^2$};
\node at (4,-2.5) {\Large$t_{12}t_{23}$};
\end{tikzpicture} 
\end{center}
\caption{Diagrammatic representation of integration over third vertex}
\label{diag: int_3}
\end{figure}
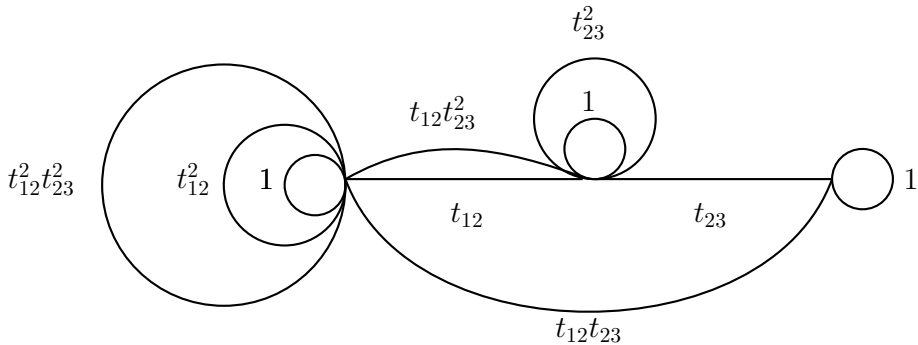

\vspace*{.06in}Finally, we integrate over the third vertex. The effect of this integration is represented by making the small circle around that vertex bigger. Drawing another circles around the first two vertices each and drawing another solid line connecting the first and second vertices. We associate a weight of 1 for the circle around the third vertex, a weight of $t_{23}^2$ for the new circle around the second vertex, a weight of $t_{12}^2t_{23}^2$ for the new circle around the first vertex and a weight of $t_{12}t_{23}^2$ for new solid line connecting the first two vertices. This is shown in Figure \ref{diag: int_3}.

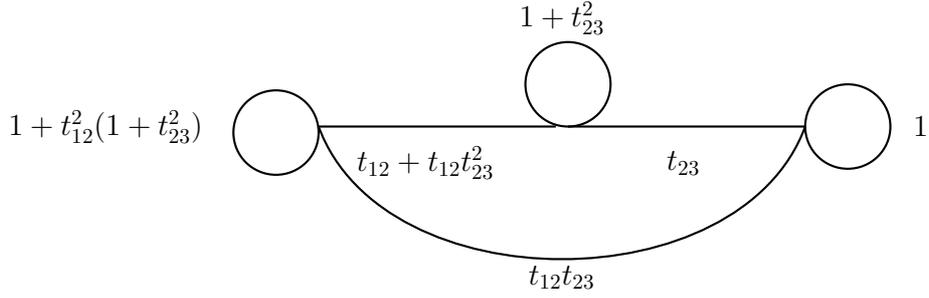
\begin{figure}
\begin{center}
\begin{tikzpicture}[scale=0.8,transform shape]
\draw[thick, black] (0,0) -- (3.9,0);
\draw[thick, black] (4.1,0) -- (8,0);
\draw[thick,black]  (-.7,-.1) circle (.7cm);
\draw[thick,black]  (8.7,0) circle (.7cm);
\draw[thick,black]  (4.1,.7) circle (.7cm);
\draw [ thick,black] (8,0) to[out=250,in=290]  (0,0);
\node at (9.9,0) {\Large$1$};
\node at (4,1.8) {\Large$1+t_{23}^2$};
\node at (-3.5,0) {\Large$1+t_{12}^2(1+t_{23}^2)$};
\draw (1.76,-.6) node {\Large $t_{12}+t_{12}t_{23}^2$};
\draw (6,-.6) node {\Large $t_{23}$};
\node at (4,-2.5) {\Large$t_{12}t_{23}$};
\end{tikzpicture} 
\end{center}
\caption{Final step for writing the Mellin amplitude}
\label{diag: int_4}
\end{figure}

Next, we replace the two lines between the first two vertices by a single line and associate a weight which is sum of the weights of previous two lines. Similarly, we replace the multiple circles at each vertex by a single circle and associate a weight which is sum of the weights of all circles initially present. After combining multiple lines and circles, Figure \ref{diag: int_3} has been redrawn in Figure \ref{diag: int_4}. 

\vspace*{.09in}To write the Mellin amplitude, 
\begin{enumerate}
\item For each initial dashed line between the vertices $u_a$ and $u_b$, we associate an integral  
$$\frac{1}{\Gamma(\inc_{ab})}\int_{0}^{\infty}dt_{ab}\;(t_{ab})^{\inc_{ab}-1}$$ 
\item For each solid line and circle in the final diagram, we include in the integrand, the corresponding weight factor raised to the power of $\mv_{ab}$ where $a$ and $b$ are the two vertices associated with the line or the circle (in which case $a=b$). 
\end{enumerate}
Following the steps described above, the Mellin amplitude for the three vertex tree can be obtained to be
\begin{eqnarray}
M(s_{ab})&=&\frac{1}{\Gamma(\inc_{12})\Gamma(\inc_{23})}
\int_0^\infty dt_{12}(t_{12})^{\inc_{12}-1}
\int_0^\infty dt_{23}(t_{23})^{\inc_{23}-1}
\nonumber\\&&
\Bigl[1+t_{12}^2(1+t_{23}^2)\Bigl]^{-\mv_{11}}
\Bigl[1+t^2_{23}\Bigl]^{-\mv_{22}}\Bigl[t_{12}(1+t^2_{23})\Bigl]^{-\mv_{12}}\Bigl[t_{12}t_{23}\Bigl]^{-\mv_{13}}\Bigl[t_{23}\Bigl]^{-\mv_{23}}
\non
\end{eqnarray}
This is same as the expression \eqref{threev7} given in the previous section. 
\subsubsection{General Rules}
\mvnote{We shall now state the general rules for writing down the Mellin amplitude for any given Feynman diagram involving scalar fields. With a little thought, we can convince ourselves that the method described above works for any diagram. This is essentially due to the reason that for deriving the Mellin amplitude of any diagram, we need to integrate over the position space vertices and introduce the Mellin variables in the same manner. Hence, the steps of integration over the vertices, for any diagram, can be captured in a diagrammatic manner as described above for three vertex tree diagram. }

\vspace*{.06in}\mvnote{We start with the skeleton and follow the steps given below for each interaction vertex, one at a time. For a general Feynman diagram there is a freedom to choose the order in which the different vertices are integrated over one by one. This procedure works for any chosen ordering. }  

\vspace*{.15in}\textbf{Diagrammatic representation of integrating over an interaction vertex}

\vspace*{.15in}At any interaction vertex on the skeleton (which has not been integrated yet), in general, there will be a small hollow circle denoting the external lines, and dashed and solid lines for the internal propagators. To represent the effect of integration over this vertex, we do the following:
\begin{enumerate}
\item Replace the small circle with a bigger circle and associate a weight 1. 
\item If this vertex is connected by a solid line (with weight $t$) to another vertex which already has a circle with some weight, draw another circle at that vertex. Associate a weight $t^{2}$ to this new circle.
\item If this vertex (that is being integrated over) is connected to another vertex with a dashed line, we replace that dashed line with a solid line and associate a weight equal to the Schwinger parameter for this internal line.

\item After the third step, if this vertex (which is being integrated over) happens to be connected to two or more vertices $\{a\}$ by solid lines with weights $\{t_{a}\}$, then we join each pair of those vertices by a solid line as well. To the new line joining vertex $a$ and $b$, we associate the weight $t_{a}t_{b}$.  

\item If any two vertices are connected by multiple lines (or any vertex has multiple circles) with each line (or circle) being associated with some weight, we replace them with a single line (or a single circle) with a weight equal to the sum of the weights of the individual lines (or circles). The final diagram should have a single line between any two vertices and a single circle at each vertex.

\item If the vertex (which is being integrated over) is only connected with internal lines but no external lines (in other words, it does not have small circle), then we do not make a bigger circle around it. However, the steps 2-5 are still applicable.
\end{enumerate}

\vspace*{.15in}\textbf{Writing the Mellin amplitude} 
\begin{enumerate}
\item For each initial dashed line between the points $a$ and $b$, we write an integral  
$$\frac{1}{\Gamma(\inc_{ab})}\int_{0}^{\infty}dt_{ab}\;(t_{ab})^{\inc_{ab}-1}$$ 
\item For each solid line between the interaction vertices $a$ and $b$ (and a circle at $a$) in the final diagram, we include in the integrand a factor equal to the corresponding weight raised to the power $\mv_{ab}$ ($\mv_{aa}$ for the circle).
\end{enumerate}

\vspace*{.09in}Although the output of this procedure is always the Mellin amplitude which is unique, the exact expression for the integrand (function of the Schwinger parameters for the internal lines) depends on the order we choose for integrating over the vertices. Also, it is not easy to show by direct evaluation of the integrals that these different integrals are in fact equal. For our purposes, we shall choose the order for integrating over the vertices that leads to the simplest integral.

\subsection{ $n$-Vertex Simple Tree}
\label{sec:simtree}
In this subsection, we consider a tree level diagram with any number of interaction vertices such that all the internal propagators are connected as a single chain. In other words, there are no branches on the skeleton as shown in Figure \ref{ntree}. We refer to this diagram as the \textit{simple $n$-vertex tree}. The examples considered in section \ref{TL} are special cases of this. 

\vspace*{.19in}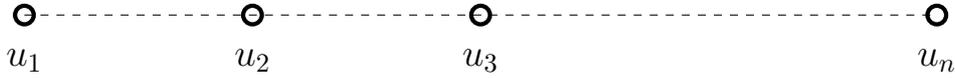
\begin{figure}[h]
\begin{center}
\begin{tikzpicture}[scale=1]
\draw [dashed]  (-6,0) -- (6,0);
\draw [ultra thick] (-6,0) circle (3.5pt);
\draw [ultra thick] (-3,0) circle (3.5pt);
\draw [ultra thick] (0,0) circle (3.5pt);
\draw [ultra thick] (6,0) circle (3.5pt);
\draw (-6,-.6) node {\Large $u_1$};
\draw (-3,-.6) node {\Large $u_2$};
\draw (0,-.6) node {\Large $u_3$};
\draw (6,-.6) node {\Large $u_n$};
\end{tikzpicture}
\end{center}
\caption{Simple tree with $n$ vertices}
\label{ntree}
\end{figure}

The position space amplitude for this diagram is given by the following integral expression
\begin{eqnarray}
I&=&\prod\limits_{a=1}^n\Biggl[\int \mathcal{D} u_a\Biggl\{\prod\limits_{i\in a}(x_a^i-u_a)^{-2\exc_a^i}\Gamma(\exc_a^i)\Biggl\}
\Biggl\{(u_a-u_{a+1})^{-2\inc_{a,a+1}}\Biggl\}\Biggl]
\non\label{fourv1}
\end{eqnarray}
\noindent
This expression can be brought to the standard form
\begin{eqnarray}
I&=&
\prod\limits_{a=1}^n\prod\limits_{b=a}^n\Biggl(
\prod\limits_{ (i, j)\in a+b}\int_{c_{ab}^{ij}-\iifty}^{c_{ab}^{ij}+\iifty}\left[d\mv_{ab}^{ij}\right]
\Gamma(\mv_{ab}^{ij})\bigl(x_{ab}^{ij}\bigl)^{-2\mv_{ab}^{ij}}\Biggl)
\prod\limits_{a=1}^{n}\prod\limits_{i\in a}
\Biggl(2\pi i\delta(\rho_a^i)
\Biggl)
M(\mv_{ab})
\label{fourv7}
\end{eqnarray}
\begin{eqnarray}
\mbox{where},\qquad\quad\rho_a^{i}\equiv\exc_a^i-\sum\limits_{j\in a}\mv_{aa}^{ij}-\sum_{\substack{b={1}\\b\not=a}}^{n}\biggl(\sum\limits_{j\in b}
\mv_{ab}^{ij}\biggl)\quad,\qquad 1\le a\le n
\label{fourv24}
\end{eqnarray}
For the standard order of integration over the position space vertices (namely, $u_1\rightarrow u_2\rightarrow \cdots \rightarrow u_n$), the Mellin amplitude $M(\mv_{ab})$ in \eqref{fourv7} is given by
\begin{eqnarray}
 M(\mv_{ab})&=&\frac{1}{\prod\limits_{a=1}^{n-1}\Gamma(\inc_{a,a+1})}
\prod\limits_{a=1}^{n-1}\Biggl\{\int_0^\infty dt_{a,a+1}\bigl(t_{a,a+1}\bigl)^{R_a-1} 
\Bigl(G_a^{n}\Bigl)^{-Q_a}\Biggl\}
\label{fourv12}
\end{eqnarray}
where,
\begin{eqnarray}
R_{a}=\inc_{a,a+1}-\sum\limits_{b=a+1}^{n}\sum
\limits_{c=1}^as_{cb}\quad,\qquad 
Q_a=\sum\limits_{b=1}^as_{ba}\quad,\qquad 1\le a\le n-1
\label{fourv13}
\end{eqnarray}
\begin{eqnarray}
G_a^b=1+t_{a,a+1}^2(1+t_{a+1,a+2}^2(\cdots\cdots+ t_{b-1,b}^2))   \quad \qquad 1\le a\le n-1,\qquad a\le b
\end{eqnarray}

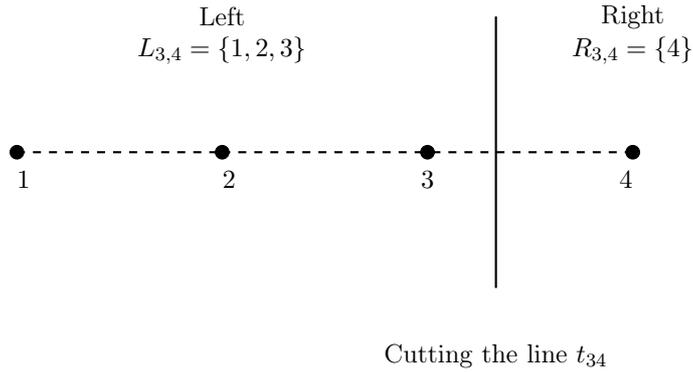
\begin{figure}[h]
\begin{center}
\begin{tikzpicture}[scale=.9,transform shape]
\filldraw[fill=black!100, draw=black]  (3,3) circle (0.10cm);
\filldraw[fill=black!100, draw=black]  (6,3) circle (0.10cm);
\filldraw[fill=black!100, draw=black]  (9,3) circle (0.10cm);
\filldraw[fill=black!100, draw=black]  (12,3) circle (0.10cm);
\node at (3.1,2.6) {$1$};
\node at (6.1,2.6) {$2$};
\node at (9,2.6) {$3$};
\node at (11.9,2.6) {$4$};
\draw[thick, black, dashed] (3,3) -- (6,3);
\draw[thick, black, dashed] (6,3) -- (9,3);
\draw[thick, black, dashed] (12,3) -- (9,3);
\draw[thick, black] (10,1) -- (10,5);
\node at (6,5) {Left};
\node at (12,5) {Right};
\node at (10,0) {Cutting the line $t_{34}$};
\node at (6,4.5) {$L_{3,4}=\{1,2,3\}$};
\node at (12,4.5) {$R_{3,4}=\{4\}$};
\end{tikzpicture}
\end{center}
\caption{Left and Right of a cut line}
\label{cut_diag}
\end{figure}

To evaluate this integral, we start with the Schwinger variable $t_{n-1,n}$ and make a coordinate transformation and a rescaling simultaneously 
\begin{eqnarray}
1+t_{n-1,n}^2=y_{n-1}\quad,\qquad 
y_{n-1}t_{n-2,n-1}^2\rightarrow t_{n-2,n-1}^2
\non
\end{eqnarray}
By making a further coordinate transformation $y_{n-1}-1\rightarrow y_{n-1}$, the integration over $y_{n-1}$ can be recognised as a beta function and we obtain
\begin{eqnarray}
 M(\mv_{ab})&=&\frac{1}{\prod\limits_{a=1}^{n-1}\Gamma(\inc_{a,a+1})}
\prod\limits_{a=1}^{n-2}\Biggl\{\int_0^\infty dt_{a,a+1}\bigl(t_{a,a+1}\bigl)^{R_a-1} 
\Bigl(G_a^{n-1}\Bigl)^{-Q_a}\Biggl\}
\nonumber\\&&
\frac{1}{2}\beta\Bigl(\frac{R_{n-1}}{2},\frac{R_{n-2}-R_{n-1}}{2}+Q_{n-1}\Bigl)
\non
\end{eqnarray}
We iteratively perform similar steps in order to integrate over the remaining Schwinger parameters and make necessary simplifications to obtain, 
\begin{eqnarray}
 M(\mv_{ab})&=&\frac{1}{\prod\limits_{a=1}^{n-1}\Gamma(\inc_{a,a+1})}
\prod\limits_{a=1}^{n-1}\frac{1}{2}\beta\Biggl(\frac{\inc_{a,a+1}-\sum\limits_{b\in L_{a,a+1}}\sum\limits_{c\in R_{a,a+1}}s_{bc}}{2}\:\;,\;\frac{D-2\gamma_{a,a+1}}{2}\Biggl)
\label{fourv22}
\end{eqnarray}
where,
\begin{eqnarray}
\sum\limits_{b\in L_{a,a+1}}\sum\limits_{c\in R_{a,a+1}}s_{bc}\equiv \sum\limits_{b=1}^a\sum\limits_{c=a+1}^{n}s_{bc}
\non\label{fourv23}
\end{eqnarray}
The $L_{a,a+1}$ and $R_{a,a+1}$ appearing in above two equations stand for Left and Right respectively. If we cut the diagram \ref{ntree} along the propagator $t_{a,a+1}$, the vertices will get divided in two sets. The set $L_{a,a+1}$ includes all the vertices which lie to to left of the cut and the set $R_{a,a+1}$ includes all the vertices which lie to the right of the cut. An example for $n=4$ is given in Figure \ref{cut_diag} in which the sets $L_{3,4}$ and $R_{3,4}$ have been shown. 

The result \eqref{fourv22} is consistent with the previous examples as the Mellin amplitude is a product over all the propagator factors (each of which is a beta function with appropriate arguments). We can again introduce dual Mellin momenta and replace the Mellin variables in the arguments of beta functions in favour of the total Mellin momenta flowing through the propagator. The propagators develop a pole when the negative of total Mellin momenta squared flowing through it becomes equal to the conformal dimension of a primary or descendant flowing through it.

\subsection{General Tree}               
 \label{TG}
\begin{figure}[h]
\begin{center}
\begin{tikzpicture}[scale=.6,transform shape]
\draw[ultra thick]  (0,0) circle (4pt);
\draw[ultra thick]  (3,3) circle (4pt);
\draw[ultra thick]  (6,0) circle (4pt);
\draw[ultra thick]  (3,6) circle (4pt);
\draw[ultra thick]  (-2,-2) circle (4pt);
\draw[ultra thick]  (-2,0) circle (4pt);
\draw[ultra thick]  (0,-2) circle (4pt);
\draw[ultra thick]  (6,-2) circle (4pt);
\draw[ultra thick]  (8,0) circle (4pt);
\draw[ultra thick]  (8,-2) circle (4pt);
\draw[ultra thick]  (5,5) circle (4pt);
\draw[ultra thick]  (1,5) circle (4pt);
\draw[ultra thick]  (5,8) circle (4pt);
\draw[ultra thick]  (1,8) circle (4pt);
\draw[thick, black, dashed] (0,0) -- (3,3);
\draw[thick, black,dashed] (3,3) -- (3,6);
\draw[thick, black,dashed] (3,3) -- (6,0);
\draw[thick, black,dashed] (0,0) -- (0,-2);
\draw[thick, black,dashed] (0,0) -- (-2,-2);
\draw[thick, black,dashed] (0,0) -- (-2,0);
\draw[thick, black,dashed] (1,5) -- (3,3);
\draw[thick, black,dashed] (5,5) -- (3,3);
\draw[thick, black,dashed] (6,0) -- (8,0);
\draw[thick, black,dashed] (6,0) -- (8,-2);
\draw[thick, black,dashed] (6,0) -- (6,-2);
\draw[thick, black,dashed] (3,6) -- (5,8);
\draw[thick, black,dashed] (1,8) -- (3,6);
\end{tikzpicture}
\end{center}
\caption{Example of a general tree level Feynman diagram (skeleton)}
\label{example_arb_tree}
\end{figure}
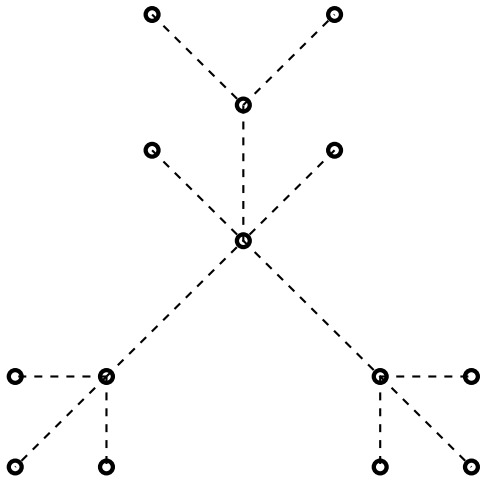

In this section, we consider a completely general tree Feynman diagram (such as the one
shown in figure \ref{example_arb_tree}) and show that the Mellin amplitude for it can be written in a simple form as product over all the internal propagator factors. 
 
The derivation of Mellin amplitude for such arbitrary tree Feynman diagrams is a graph theoretic exercise and does not shed any light on the physical significance of the result itself. Hence, in this section, we shall only state the final result and discuss it's physical significance. The details of the derivation have been presented in the appendix \ref{general_tree_proof}.

\vspace*{.06in}From the diagrammatic rules given in section \ref{sec: diag_rules}, we know that the amplitude can be written in the following form
\begin{equation}
M\left(\{s_{ab}\}\right)=\prod\left[ \int_{0}^{\infty}dt_{ab}\frac{(t_{ab})^{\gamma_{ab}-1}}{\Gamma(\gamma_{ab})}\right]F\Bigl(\{t_{ab}\},\{s_{ab}\}\Bigl)   
 \label{Mellin_general}
\end{equation}

The product runs over all the internal lines\footnote{Since we are considering a general tree which may have brances, it is not necessary that neighbouring vertices will always be labelled with consecutive integers.}. $F$ is function of the Schwinger parameters and the Mellin variables.

\vspace*{.06in}The function $F$ depends on the order of integration of the position space vertices and, in general, is a very complicated function of the Schwinger parameters. It turns out that for the tree diagrams, it is possible to make a choice for the order in which the vertices are integrated over such that the  integral \eqref{Mellin_general} can be performed easily. This has been described in detail in the appendix \ref{general_tree_proof}. With such a choice, the function $F$ can be expressed as
\begin{eqnarray}
F\;\;=\prod\limits_{\mbox{all propagators}} (t_{ab})^{-P_{ab}}(A_{ab})^{-Q_{ab}}  \non \label{laddgt1}
\end{eqnarray}
where,
\begin{eqnarray}
\nonumber && P_{ab}=\sum\limits_{c\in L_{ab}}\sum\limits_{d\in R_{ab}}s_{cd} \qquad;\qquad Q_{ab}=\sum_{\substack{c\in L_{ab}\\c\not=a}}\sum_{\substack{d\in \tilde L_{ab}\\d\not=a}}s_{cd}\quad+\sum\limits_{d\in L_{ab}}s_{ad}               \label{laddgt3}
 \end{eqnarray}
The term involving the double sum in $Q_{ab}$ is absent if there is no branching at the vertex $a$ in the skeleton. The tilde in one of the $L$ in this double sum denotes the fact that we should not include terms of the type $s_{cd}$ where $c$ and $d$ are on the same branch in the set $L_{ab}$.

\vspace*{.06in}To define the function $A_{ab}$, we shall need a reference vertex which can be chosen freely from any one of the end vertices (a vertex with only one dashed line attached to it) on the skeleton. Let that vertex be $\mathcal{P}$ (see figure \ref{vertex_order} and the related discussion in appendix \ref{general_tree_proof}), then
\begin{eqnarray}
 &&A_{ab}\equiv1+t_{ab}^{2}\left(1+t_{bc}^{2}\left(1+...(1+t_{o\mathcal{P}}^{2})...\right)\right) 
 \non   \label{laddgt2}
\end{eqnarray}

\vspace*{.06in}$b$,$c$,...,$o$ are all on the shortest continuous route from $a$ to the reference vertex $\mathcal{P}$.

\vspace*{.06in}The final result, after integrating over all the Schwinger parameters in \eqref{Mellin_general}, turns out to be a product of beta functions with one beta function for each internal propagator. The arguments of beta functions involve the Left and Right part of the propagator as in the case of simple tree in previous subsection. Since there may be branches in our tree, we need to specify what Left and Right of a cut line mean in this context. As a rule, we refer to the part of the diagram (after the cut) having the reference vertex $\mathcal{P}$ as the Right. With this, we can write the Mellin amplitude for a completely general tree as
\begin{equation}
M\left(\{s_{ab}\}\right)=\prod \frac{1}{2\Gamma(\gamma_{ab})}\beta\left(\frac{\gamma_{ab}-\sum\limits_{c\in L_{ab}}\sum\limits_{d\in R_{ab}}s_{cd}}{2} \ , \ \frac{D}{2}-\gamma_{ab}\right)       
\label{chap6eqn24}
\end{equation}
The product is over all the internal propagators of the diagram.


\vspace*{.06in}The physical interpretation of the above result becomes clear if we again consider the dual Mellin momenta. As before, we denote the Mellin momentum flowing into the diagram through the external vertex $x_a^i$ by $k_a^i$. We first note that the constraints on the Mellin variables are automatically satisfied if the total Mellin momentum is conserved. The constraint satisfied by the Mellin variables is
\be
\sum_{\substack{b}}\biggl(\sum\limits_{j\in b}
\mv_{ab}^{ij}\biggl)=\exc_a^i\qquad \implies\qquad k_a^i\cdot\biggl(\sum_{\substack{b}}\sum\limits_{j\in b}
k_b^j\biggl)=0
\non\label{mellin_momenta}
\ee
where, we have used $\Delta_a^i= -(k_a^i)^2$.

\vspace*{.06in}The simplest way to satisfy the above equation is by demanding that the total Mellin momenta is conserved, namely $\sum\limits_{\substack{b}}\sum\limits_{j\in b}
k_b^j=0$. 

Now, the full Mellin momentum propagating through an internal propagator, joining the vertices $a$ and $b$, is 
\begin{eqnarray}
k=\sum_{c\in L_{ab}}\sum_{i \in c}k^i_c=
-\sum_{c\in R_{ab}}\sum_{i \in c}k^i_c
\non
\end{eqnarray}
and the kinematical variable in the propagator is 
\begin{eqnarray}
\sum\limits_{c\in L_{ab}}\sum\limits_{d\in R_{ab}}s_{cd}=\sum\limits_{c\in L_{ab}}\sum\limits_{d\in R_{ab}}\sum_{i\in c}\sum_{j \in d}s_{cd}^{ij}=\left(\sum_{c\in L_{ab}}\sum_{i \in c}k^i_c\right)\cdot\left(\sum_{d\in R_{ab}}\sum_{j \in d}k^j_d\right)=-k^2\non
\end{eqnarray}

\vspace*{.06in}In terms of the Mellin momenta, the expression for the propagator can be written as
\begin{equation}
\frac{1}{2\Gamma(\gamma_{ab})}\beta\left(\frac{1}{2}\left(\gamma_{ab}+k^{2}\right), \frac{D}{2}-\gamma_{ab}\right)=\frac{1}{\Gamma(\gamma_{ab})}\sum_{n=0}^{\infty}\:\frac{\Gamma\left(\gamma_{ab}-\frac{D}{2}+1+n\right)}{n!\Gamma\left(\gamma_{ab}-\frac{D}{2}+1\right)}\frac{1}{k^{2}+\gamma_{ab}+2n}       
\non    
\end{equation}

\vspace*{.06in}This shows that the total Mellin momentum (squared) flowing through the propagator has poles at $-\gamma_{ab}-2n$. These correspond to the propagation of a primary field and the corresponding descendants \mvnote{(see footnote 3)}. The above sum representation of the propagator is analogous to the K{\"a}hl{\'e}n Lehmann spectral representation in ordinary quantum field theories. 

\vspace*{0.06in}The Feynman rules for tree level Feynman diagrams in perturbative CFT for scalar fields is now obvious. The propagator for any internal line is given by \eqref{chap6eqn24} and we simply multiply all the propagator factors of the diagram. 

\vspace*{.06in}We would like to wrap up this discussion with a brief recapitulation of the most important points we have learnt so far. Mellin space provides a manifest conformally invariant representation for correlation functions in a CFT. At tree level, there exist a set of Mellin space Feynman rules that can be associated with Feynman diagrams involving  scalar operators. Some linear combinations of the Mellin variables that appear in the propagators can be interpreted as Mandelstam variables constructed out of the (hypothetical) external Mellin momenta flowing into the diagram. The invariance of the amplitude under special conformal transformation allows for a statement of conservation of Mellin momentum. All the Mellin variables (or equivalently all the Mandelstam variables) are not independent and the number of independent Mellin variables is equal to the number of independent cross ratios between the external vertices in the diagram. Mellin space also allows a spectral representation for the correlation functions as any propagator in the diagram has a discrete infinite set of poles corresponding to the exchanged primary field and its descendants.

\section{One-Loop Feynman Diagram}
\label{mellin_loop}
After deriving the Mellin space Feynman rules for tree level diagrams, the next step is to consider loop diagrams. We have not yet been able to derive the Feynman rules for loop diagrams. In this section, we shall content ourselves with the expression for the one-loop Mellin amplitude as an integral over the internal Schwinger parameters. It may be possible to derive the loop Feynman rules in Mellin space using an approach similar to the one presented in appendix \ref{n vertex mellin} which treats the $n$-vertex simple tree in a way different from what we have already seen in section \ref{sec:diag}. 

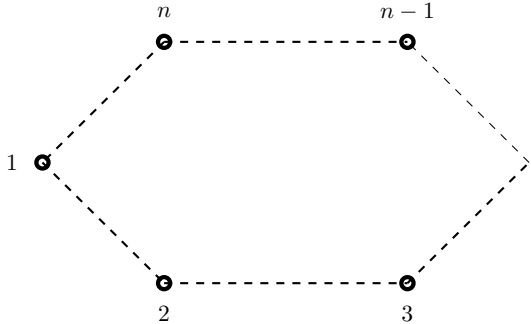
\begin{figure}[h]
\begin{center}
\begin{tikzpicture}[scale=.8,transform shape]
\node at (3.5,-5) {$1$};
\draw[ultra thick]  (4,-5) circle (0.10cm);
\draw[thick, dashed] (4,-5) -- (6,-7);
\node at (6,-7.5) {$2$};
\draw[ultra thick]  (6,-7) circle (0.10cm);
\draw[thick, dashed] (6,-7) -- (10,-7);
\node at (10,-7.5) {$3$};
\draw[ultra thick]  (10,-7) circle (0.10cm);
\draw[thick, dashed] (10,-7) -- (12,-5);
\draw[dashed, black] (12,-5)--(10,-3)  ;
\node at (10,-2.5) {$n-1$};
\draw[ultra thick]  (10,-3) circle (0.10cm);
\draw[thick, dashed] (6,-3) -- (10,-3);
\node at (6,-2.5) {$n$};
\draw[ultra thick]  (6,-3) circle (0.10cm);
\draw[thick, dashed] (6,-3) -- (4,-5);
\end{tikzpicture}
\end{center}
\caption{One loop diagram with $n$ internal vertices}
\label{diag:ngon1}
\end{figure}

\vspace*{.06in}The position space amplitude for the loop Feynman diagram in figure \eqref{diag:ngon1} is given by 
\\
\begin{eqnarray}
I&=&\prod\limits_{a=1}^n\Biggl[\int \mathcal{D} u_a\prod\limits_{i\in a}\left\{(x_a^i-u_a)^{-2\exc_a^i}\Gamma(\exc_a^i)\right\}
\Biggl]\prod\limits_{b=1}^n(u_b-u_{b+1})^{-2\inc_{b,b+1}}\label{ngon1}
\end{eqnarray}
where $n+1\equiv 1$.

\subsection{One-Loop Mellin Amplitude}
The Mellin amplitude for the $n$-vertex one-loop diagram can be derived using the position space amplitude \eqref{ngon1} by following the same procedure as in the previous sections for tree diagrams. The Mellin amplitude is written as integral over the Schwinger parameters and the integrand depends upon the order in which we perform the integration over the interaction vertices in position space. For the cyclic order of integration ($u_1\rightarrow u_2\rightarrow \cdots \rightarrow u_n$), the Mellin amplitude turns out to be
\begin{eqnarray}
M(\mv_{ab})&=&
\prod\limits_{a=1}^n\left[\int_0^\infty dt_{a,a+1}\frac{(t_{a,a+1})^{\inc_{a,a+1}-1}}{\Gamma(\gamma_{a,a+1})} \right]
\prod_{a=1}^{n-1}\prod_{b=a}^n\Bigl(\tilde H_a^b+\tilde K_a\tilde K_b\Bigl)^{-\mv_{ab}}
\label{eqnloop2}
\end{eqnarray}
where,
\begin{eqnarray}
\tilde H_a^b&\equiv&t_{a, a+1}\cdots t_{b-1,b} G_b^{n-1} \quad \quad \quad \quad \quad \quad \quad \quad \quad \quad \quad \quad 1\le  a<b\le n-1
\nonumber\\
\tilde H_a^a&\equiv& G_a^{n-1}\qquad \qquad \quad \quad \quad \quad \qquad \quad \quad \quad \quad \quad \quad \quad \quad 1\le  a\le n-1
\nonumber\\
\tilde H_a^n&\equiv&0 \qquad \qquad  \quad \quad \quad \quad \quad \qquad \qquad  \quad \quad \quad \quad \quad \quad \quad  1\le  a\le n \nonumber \\
\tilde K_a &\equiv&\Big(t_{1n} t_{12} \cdots t_{a-1,a} G_a^{n-1}+t_{a,a+1}\cdots t_{n-1,n}\Big)\quad \quad \quad   1\le a \le n-1
\nonumber\\
\tilde K_n&=&1 \nonumber 
\end{eqnarray}
\vspace*{0.06in} We have defined $G_{a}^{b}$ in eq. \ref{fourv14}.

\subsection{A Consistency Check}

\vspace*{0.06in} For the consistency of the diagrammatic algorithm (which works for any Feynman diagram) given in section \ref{sec: diag_rules}, the one loop amplitude considered above should reduce to the $n$-vertex simple tree amplitude when we cut a propagator of the loop.\footnote{Operationally, this can be done by ``removing'' the integration over the corresponding Schwinger parameter.} This indeed turns out to be the case which is a sanity check for the result \eqref{eqnloop2}.

\vspace*{0.06in} It turns out that removing different propagators corresponds to the tree result written in different forms. These different forms correspond to choosing different order of integration for the position space vertices of the tree. For example, setting $t_{a,a+1}$ to zero gives the corresponding result for $n$ vertex simple tree for which the integration has been performed in the order $u_{a+1}\rightarrow u_{a+2}\rightarrow \cdots \rightarrow u_n\rightarrow u_{1}\rightarrow \cdots \rightarrow u_a$. As a special case, the limit $t_{1n}\rightarrow 0$ gives the result for the standard order of integration ($u_1\rightarrow \cdots\rightarrow u_n $) over the position space vertices.

\subsection{Special Case: Loop With 3 Internal Vertices}
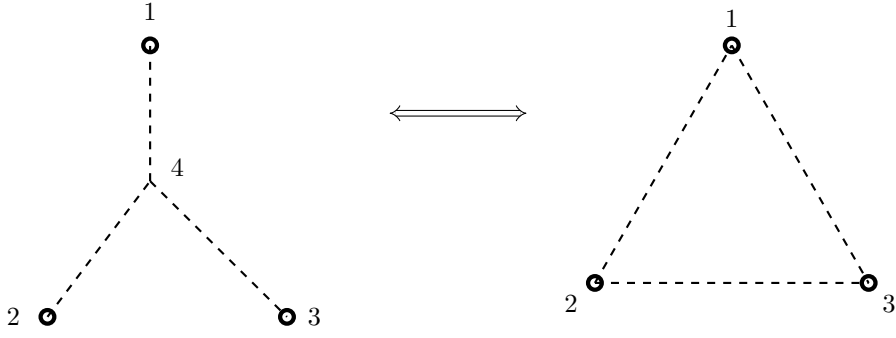
\begin{figure}[h]
\begin{center}
\begin{tikzpicture}[scale=.9,transform shape]
\draw[ultra thick]  (-1,2) circle (0.10cm);
\draw[thick, dashed] (-1,0) -- (-1,2);
\node at (-.6,0.2) {$4$};
\node at (-1,2.5) {$1$};
\draw[ultra thick]  (-2.5,-2) circle (0.10cm);
\draw[thick, dashed] (-2.5,-2) -- (-1,0);
\node at (-3,-2) {$2$};
\draw[ultra thick]  (1,-2) circle (0.10cm);
\draw[thick, dashed] (-1,0) -- (1,-2);
\node at (1.4,-2) {$3$};
\draw[thick, dashed] (5.5,-1.5) -- (7.5,2);
\node at (7.5,2.4) {$1$};
\draw[ultra thick]  (7.5,2) circle (0.10cm);
\draw[thick, dashed] (7.5,2) -- (9.5,-1.5);
\node at (5.15,-1.8) {$2$};
\draw[ultra thick]  (5.5,-1.5) circle (0.10cm);
\draw[ultra thick]  (9.5,-1.5) circle (0.10cm);
\draw[thick, dashed] (5.5,-1.5) -- (9.5,-1.5);
\draw[implies-implies,double equal sign distance] (2.5,1) -- (4.5,1);
\node at (9.8,-1.8) {$3$};

\end{tikzpicture}
\end{center}
\caption{Skeleton of the ``star'' and the ``delta''}
\label{diag:stdel}
\end{figure}

The conformal Mellin amplitude of one loop diagram with 3 internal vertices ( ``delta'' diagram) can be exactly evaluated in terms of a tree amplitude (``star'' tree diagram). This happens due to the standard ``star-delta'' relation (see figure \ref{diag:stdel}) in an analogy with a similar result in electrical circuits. 

The position space expression for the star diagram is
\begin{eqnarray}
I_{star}&=&\int \mathcal{D}u_4 \prod\limits_{a=1}^{3}\Biggl[\int\mathcal{D}u_a
\biggl\{\prod\limits_{i\in a}(x^i_a-u_a)^{-2\exc^i_a}\Gamma(\exc^i_a)\biggl\}(u_a-u_4)^{-2\inc'_{a,4}}\Biggl]
\non
\end{eqnarray}
To show the equivalence with the 3 vertex loop, we need to perform the integration over the central vertex $u_4$. For this vertex, we perform the standard algebraic steps as in section \ref{T1} and obtain
\begin{eqnarray}
&&\hspace*{-.1in}I_{star}=
\non\\
&&
\prod\limits_{a=1}^{3}\Bigg\{\int \mathcal{D}u_a\prod_{i\in a}(x^i_a-u_a)^{-2\exc^i_a}\Gamma(\exc^i_a)\Bigg\}
\prod\limits_{a=1}^{3}\left(\int_{0}^\infty dt_{a,4}\;\frac{(t_{a,4})^{\inc'_{a,4}-1}}{\Gamma(\inc'_{a,4})}\right)
\nonumber\\
&&
\left(\prod\limits_{a=1}^3\int [d\mv_a]\Gamma(\mv_a)
\right)\Bigl\{t_{14}t_{24}(u_1-u_2)\Bigl\}^{-2\mv_3}\Bigl\{t_{14}t_{34}(u_1-u_3)\Bigl\}^{-2\mv_{2}}\Bigl\{t_{24}t_{34}(u_2-u_3)\Bigl\}^{-2\mv_1}
\nonumber
\end{eqnarray} 
Integration over the Schwinger parameters $t_{a,4}$ give 3 delta functions. We can thus perform the 3 integrals over $\mv_a$. The resulting expression is proportional to the 3 vertex loop amplitude in position space, i.e. 
\begin{eqnarray}
I_{star}&=&\frac{\Gamma(\inc_{12})\Gamma(\inc_{23})\Gamma(\inc_{13})}{\prod\limits_{a=1}^{3}\Gamma(\inc'_{a,4})}
\prod\limits_{a=1}^{3}\Bigg\{\int \mathcal{D}u_a\prod_{i\in a}(x^i_a-u_a)^{-2\exc^i_a}\Gamma(\exc^i_a)\Bigg\}
\nonumber\\
&&
\times\ (u_1-u_2)^{-2\inc_{12}}(u_1-u_3)^{-2\inc_{13}}(u_2-u_3)^{-2\inc_{23}}
\nonumber
\end{eqnarray}
where,
\begin{eqnarray}
  \inc'_{14}=\inc_{12}+\inc_{13}\qquad,\quad  \inc'_{24}=\inc_{12}+\inc_{23}\qquad,\quad \inc'_{34}=\inc_{13}+\inc_{23}
  \non
\end{eqnarray}
Thus, if we represent the internal propagator of 3 vertex loop by $\inc_{ab}$ and those of the star diagram by $\inc'_{a4}$, then the above relation says
\begin{eqnarray} 
I_{star}(\mv_{ab},\inc'_{ab})=\frac{\Gamma(\inc_{12})\Gamma(\inc_{23})\Gamma(\inc_{13})}{\Gamma(\inc'_{14})\Gamma(\inc'_{24})\Gamma(\inc'_{34})}\;\;\times\;\; I_{delta}(\mv_{ab},\inc_{ab})
\non
\end{eqnarray}
The star diagram is a tree diagram and its Mellin space amplitude can be easily written down using the Feynman rules given in the previous sections. Thus, we find the 3 vertex loop amplitude in Mellin space to be
\begin{eqnarray}
&&\hspace*{-.28in}I_{delta}
=\ \frac{1}{8\Gamma(\inc_{12})\Gamma(\inc_{23})
\Gamma(\inc_{13})}\beta\left(\frac{\inc_{12}+\inc_{13}-\mv_{12}-\mv_{13}}{2},\frac{D}{2}-\inc_{12}-\inc_{13}\right)
\nonumber\\
&&\hspace*{-.0in}\beta\left(\frac{\inc_{12}+\inc_{23}-\mv_{12}-\mv_{23}}{2},\frac{D}{2}-\inc_{12}-\inc_{23}\right)\beta\left(\frac{\inc_{13}+\inc_{23}-\mv_{13}-\mv_{23}}{2},\frac{D}{2}-\inc_{13}-\inc_{23}\right)
\nonumber
\end{eqnarray}

\section{Non-Conformal Mellin Amplitudes}           
\label{OA}
In this section, we revisit some of the tree level Feynman diagrams we have been considering so far. However, this time we relax the conformality conditions imposed on them.  The motivation for defining these ``non-conformal Mellin amplitudes'' comes from noting that exactly marginal deformations of CFTs are rare (generally arising only in some special supersymmteric gauge theories). 

\vspace*{0.06in} We have been considering CFTs whose Lagrangian descriptions are in terms of scalar fields. Since the couplings generically run with the energy scale, the beta function is non-zero and conformal invariance is broken.  Thus, even a classically marginal perturbation generally breaks conformal invariance once quantum effects are included\footnote{We would like to thank R. Loganayagam for discussions on this issue.}. We thus study these non-conformal Mellin amplitudes by considering a generic scalar perturbation around a free CFT that may not preserve any of the conformal or scale symmetry. At an operational level, we relax the conformality conditions that we have been imposing at each interaction vertex.



\subsection{Some Examples }
\label{some_examples}
\vspace*{0.06in}As a concrete example, we consider the simple $n$-vertex tree of figure \ref{ntree}. The conformal Mellin amplitude for this diagram was given in \eqref{fourv12}. If we do not impose the conformality conditions, then instead of \eqref{fourv7}, we obtain
 \begin{eqnarray}
I&=&
\prod\limits_{a=1}^n\prod\limits_{b=a}^n\Biggl(
\prod\limits_{ (i, j)\in a+b}\int_{-\iifty}^{\iifty}\left[d\mv_{ab}^{ij}\right]
\Gamma(\mv_{ab}^{ij})\bigl(x_{ab}^{ij}\bigl)^{-2\mv_{ab}^{ij}}\Biggl)
\tilde M(\mv_{ab})
\nonumber
\end{eqnarray}
where,
 \begin{eqnarray}
\tilde M(s_{ab})&=&\prod\limits_{a=1}^{n}\Biggl(\prod\limits_{i\in a}\int_0^\infty d\alpha_a^i(\alpha_a^i)^{\rho_a^i-1}\Biggl)
 \prod\limits_{a=1}^{n-1}\Biggl\{\int_0^\infty \frac{dt_{a,a+1}}{\Gamma(\gamma_{a,a+1})}\bigl(t_{a,a+1}\bigl)^{R_a-1} 
\Bigl(G_a^{n}\Bigl)^{-Q_a}\Biggl\}
\label{Mellin_off_shell1}
\nonumber\\
&&\times\prod\limits_{a=1}^{n}\left\{\sum\limits_{b=1}^{a-1}\;\left(H_b^{\;a}\sum\limits_{i\in b}\alpha_b^i\right)+
\sum\limits_{b=a}^{n}\;\left(H_a^{\;b}\sum\limits_{i\in b}\alpha_b^i\right)\right\}^{-\lambda_a}
\non\\
&\equiv&\delta\left(\sum\limits_{a=1}^n(\lambda_a-\rho_a)\right)M(s_{ab})
\end{eqnarray}
where $\rho_a^i$ is defined in equation \eqref{fourv24} and 
\begin{eqnarray}
\nonumber G_a^c&=&1+t_{a,a+1}^2(1+t_{a+1,a+2}^2(\cdots\cdots+ t_{c-1,c}^2))   \quad \qquad 1\le a\le n-1\qquad \\
\nonumber H_a^{\;b}&=& t_{a,a+1}t_{a+1,a+2}\cdots t_{b-1,b}G_b^{\;n}
\qquad,\qquad H_a^{\;a}= G_a^{\;n}
\non\\
\lambda_a&=& D-\sum\limits_{i\in a}\Delta_a^i-\inc_{a-1,a}-\inc_{a,a+1}\quad,\qquad 1\le a\le n
\nonumber\\
\rho_a&=&\sum\limits_{i\in a}\rho_a^i \quad,\qquad 1\le a\le n
\non
\end{eqnarray}
It is a simple exercise to extract the overall delta function from the expression of $\tilde M(s_{ab})$ as we have done in equation \eqref{Mellin_off_shell1}. We shall take $M(s_{ab})$ in \eqref{Mellin_off_shell1} to be the definition of the ``non-conformal Mellin amplitude'' for $n$-vertex simple tree\footnote{We are using the same symbol $M(s_{ab})$ to denote both conformal as well as non-conformal Mellin amplitudes. However, the distinction should be clear from the context.}. 

\vspace*{.07in}One crucial limitation of this treatment that should be noted here is that the delta function in \eqref{Mellin_off_shell1} is graph dependent and consequently the definition of $M(s_{ab})$ is also graph dependent. Therefore, although we can calculate $M(s_{ab})$ for individual diagrams, that is not exactly equal to doing perturbation theory in Mellin space. The delta function emerges in this context from the fact that the position space integrals still scale in a given way, although this scaling property depends on the particular graph being considered and may not refer to any symmetry of the theory itself.

\vspace*{.07in}The conformality condition amounts to setting all $\lambda_a$ equal to zero. However, we  work with non zero $\lambda_a$. We have not been able to obtain any simple set of Feynman rules from the general expression given in \eqref{Mellin_off_shell1}. We shall content ourselves with the special cases of $n=1$ and $n=2$ which give us some interesting results.


\subsubsection{Contact Interaction}
For a single vertex (i.e. $n=1$), the expression \eqref{Mellin_off_shell1} gives
\begin{eqnarray}
\tilde M(s^{ij})=\prod\limits_{i=1}^N\Biggl(\int_0^\infty d\alpha^i(\alpha^i)^{\rho^i-1}\Biggl)\Bigl(\sum\limits_{i=1}^N\alpha^i\Bigl)^{\sum\limits_{i=1}^N\Delta^i-D}
\non
\end{eqnarray}
All symbols have their usual meaning as used previously. In the conformal case, the expression above just gives delta function constraints on the Mellin variables. We now evaluate this in the non conformal case. For this, we insert the partition of unity
\begin{eqnarray}
1=\int_0^\infty dq\;\; \delta \left(q-\sum\limits_{i=1}^N\alpha^i\right)
\non
\end{eqnarray}
in the above integral, make the coordinate transformations $\alpha^i=q \;y^i $ and use the identity
\be
\prod\limits_{i=1}^N\int_0^\infty dx^i \ (x^i)^{\rho^i-1}\delta \left(1-\sum\limits_{i=1}^Nx^i\right)=\frac{\prod\limits_{i=1}^N\Gamma(\rho^i)}{\Gamma\left(\sum\limits_{i=1}^N\rho^i\right)}
\label{generalized_beta}
\ee
Using the expression for $\rho_i$, we finally obtain,
\begin{eqnarray}
M(\mv^{ij})&=&
\frac{\prod\limits_{i=1}^N\Gamma\left(\Delta^i-\sum\limits_{\substack{j={1}\\j\not=i}}^N
\mv^{ij}\right)}{\Gamma\left(D-\sum\limits_{i=1}^N\Delta^i\right)}
\label{offshell_1}
\end{eqnarray}
We have used the constraint arising from the overall delta function (see the definition \eqref{Mellin_off_shell1}) to simplify the arguments of the Gamma function in denominator. 

\subsubsection{Tree With One Internal Propagator}
\label{2vertex_offshell}
We next consider the tree diagram with one internal propagator (i.e. $n=2$). For $n=2$, the expression \eqref{Mellin_off_shell1} reduces to 
\begin{eqnarray}
\tilde M(\mv_{ab})&=&\frac{1}{\Gamma(\inc)}\prod\limits_{a=1}^2\prod\limits_{i\in a}\Biggl(\int_0^\infty d\alpha_a^i(\alpha_a^i)^{\rho_a^i-1}\Biggl)
\int_0^\infty dt\;(t)^{\inc-s_{12}-1}\bigl(1+t^2\bigl)^{-s_{11}}
\nonumber\\
&&\left(\Bigl(1+t^2\Bigl)\sum_{i\in1}\alpha_1^i+t\sum_{i\in2}
\alpha_2^i\right)^{-\lambda_1}
\Bigl(\sum_{i\in2}\alpha_2^i+t\sum_{i\in1}\alpha_1^i\Bigl)
^{-\lambda_2}
\label{starting_expression}
\end{eqnarray}
where, we have relabelled $t_{12}\rightarrow t$ and $\gamma_{12}\rightarrow \gamma$ to match with our notation in section \ref{TL}. After some manipulations (see Appendix \ref{appendix: off_shell}), the non-conformal Mellin amplitude can be extracted to be,
\begin{eqnarray}
&&\hspace*{-.08in}M(\mv_{ab})=\non\\
&&{_3F_{2}}\left(\gamma-\frac{D}{2}+\lambda_{1}+\lambda_{2}\;,\;\frac{R_1+\rho_1-\lambda_1}{2}\;,\;\frac{R_1+\rho_2-\lambda_2}{2}\;;\;\frac{R_1+\rho_1+\lambda_1}{2},\frac{R_1+\rho_2+\lambda_2}{2}\;;\;1\right)
\nonumber\\
&&
\times\ \frac{1}{2\Gamma(\inc)}\prod\limits_{a=1}^{2}
\left[\prod\limits_{i\in a}\Gamma(\rho_a^i)\right]
\left[\frac{\Gamma\left(\frac{R_1+\rho_1-\lambda_1}{2}\right)\Gamma\left(\frac{R_1+\rho_2-\lambda_2}{2}\right)}{\Gamma\left(\frac{R_1+\rho_1+\lambda_1}{2}\right)\Gamma\left(\frac{R_1+\rho_2+\lambda_2}{2}\right)}\right]
\label{offshellpropagator}
\end{eqnarray}
where, $R_{1}=\gamma-s_{12}$ and $\lambda_{a}=D-\gamma-\sum\limits_{i\in a}\Delta_{a}^{i}$.

 We now look at the pole structure and the conformal limit of the expression \eqref{offshellpropagator}.

\subsubsection*{Pole Structure}

The poles of the amplitude \eqref{offshellpropagator} occur when the arguments of the gamma functions in the numerator in second line are zero or negative integers (${_3}F_2$ does not give rise to any pole). Thus, as a function of the Mellin variables, the poles of the amplitude lie at
\begin{eqnarray}
\frac{R_1+\rho_1-\lambda_1}{2}=-n\qquad\implies\qquad \mv_{11}+\mv_{12}=\sum\limits_{i\in 1}\Delta_a^i+\gamma-\frac{D}{2}+2n
\non   \label{noncpoles1}
\end{eqnarray}
and,
\begin{eqnarray}
\frac{R_1+\rho_2-\lambda_2}{2}=-n'\qquad\implies\qquad \mv_{22}+\mv_{12}=\sum\limits_{i\in 2}\Delta_a^i+\gamma-\frac{D}{2}+2n' 
\non  \label{noncpoles2}
\end{eqnarray}
where, $n,n'$ are zero or arbitrary positive integers, i.e. $0,1,2,\cdots$.

\vspace*{.07in}This shows that in the non-conformal case there are two sets of poles, which in the conformal limit $\lambda_a \rightarrow 0$, coalesce to give the one set of poles we had earlier. It would be interesting to find the physical interpretation, if any, of these two sets of poles.

\subsubsection*{Conformal Limit}           
\label{conflim}

In the conformal limit, we have $\lambda_1,\,\lambda_2 \rightarrow 0$. To impose this limit, we first take $\lambda_1\rightarrow 0$ keeping $\lambda_2$ fixed and non-zero. In this limit, two of the arguments of the $_3F_2$ hypergeometric function in \eqref{offshellpropagator} will become identical and it will thus reduce to a $_2F_1$ hypergeometric function
\begin{eqnarray}
\tilde M(\mv_{ab})&=&
\frac{1}{2\Gamma(\inc)}\prod\limits_{a=1}^{2}
\left[\prod\limits_{i\in a}\Gamma(\rho_a^i)\right]
\left[\frac{\Gamma\left(\frac{R_1+\rho_2-\lambda_2}{2}\right)}{\Gamma\left(\frac{R_1+\rho_2+\lambda_2}{2}\right)}\right]
\delta\Bigl({\rho_1+\rho_2-\lambda_2}\Bigl)
\nonumber\\
&&{_2F_{1}}\left(\gamma-\frac{D}{2}+\lambda_2\;,\;\frac{R_1+\rho_2-\lambda_2}{2}\;;\;\frac{R_1+\rho_2+\lambda_2}{2}\;;\;1\right)
\nonumber
\end{eqnarray}
Now using the Gauss identity \eqref{iden5}, we obtain after some simplification
\begin{eqnarray}
\tilde M(\mv_{ab})&=&
\frac{1}{2\Gamma(\inc)}\prod\limits_{a=1}^{2}
\left[\prod\limits_{i\in a}\Gamma(\rho_a^i)\right]
\left[\frac{\delta\Bigl({\rho_1+\rho_2-\lambda_2}\Bigl)}{\Gamma\left(\lambda_2\right)}\right]
\beta\left(\frac{R_1+\rho_2-\lambda_2}{2}\;,\;\frac{D}{2}-\inc\right)
\non
\end{eqnarray}
If we now take the limit $\lambda_2\rightarrow 0$, we recover the Mellin amplitude \eqref{twov10} for the conformal case along with the delta function constraints \eqref{twov6} on the Mellin variables.

\subsection{Scale Invariant Amplitudes and Off-Shell Interpretation}
\label{6.3}
In usual QFTs, we often consider correlation functions in which the external legs are off-shell. One puts these external legs on-shell via the LSZ procedure. It turns out that we can define analogous ``off-shell'' objects for conformal field theories in Mellin space as well. 
 
\vspace*{.07in} For this purpose, we consider position space correlation functions with scale covariance (as in a theory with scale symmetry) \footnote{ In our notations, this would be equivalent to setting $\sum\limits_{a}\lambda_{a}
=0$ although individual $\lambda_{a}$ need not be equal to $0$.} although they need not have the full conformal covariance. We expect any physically interesting scale invariant theory to be conformally invariant as well (see \cite{Nakayama} and the references therein). However, it is still interesting to consider this case, since as we shall argue below, the corresponding ``Mellin amplitudes'' seem to be ``off-shell'' quantities that reduce to the ``on-shell'' Mellin amplitudes\footnote{$(k_a^i)^2=-\Delta_a^i$ being the on-shell condition.} of conformally invariant theories through an LSZ like procedure.

\vspace*{.07in}We can imagine extending the definition of the Mellin amplitude to scale invariant theories in the following manner
\begin{eqnarray}
\nonumber A\left(x^{i}\right)=\prod_{ i<j}\left(\int_{-i\infty}^{i\infty}\frac{ds^{ij}}{2\pi i}\Gamma(s^{ij})(x^{i}-x^{j})^{-2s^{ij}}\right)\delta\left(\sum_{i}\Delta^{i}-\sum_{i\neq j}\sum_{j}s^{ij}\right){M}\left(s^{ij}\right) \non
\end{eqnarray}
 As opposed to the $N$ (number of external lines) delta function constraints for the conformal amplitude \eqref{Mellin1}, in this case we only have one overall constraint on the Mellin variables resulting from the covariance under scale transformations. Therefore, the number of Mellin variables in this case is only $N(N-1)/2-1$ which is also the correct number of independent kinematical variables in a scale invariant theory. 

\vspace*{0.06in}For this case also, we can introduce the dual Mellin momenta in exactly the same manner as before (see section \ref{sec:mellin}). In terms of these dual momenta, the overall delta function constraint 
translates to the condition $\sum_{i}(k^{i})^{2}=-\sum_{i} \Delta^{i}$ which is weaker than the conformal case $(k^{i})^{2}=-\Delta^{i}$. However, we can still demand the conservation of these dual momenta, i.e. $\sum_{i}k^{i}=0$\footnote{This is not possible when the scale covariance is also absent.}.

\vspace*{0.06in}The contact interaction diagram has only one interaction vertex and ensuring scale invariance automatically ensures conformal invariance as well. However, for more than one interaction vertices, there is a difference between the scale and conformally invariant Mellin amplitudes and the expressions for the former can be obtained by imposing $\sum_{a}\lambda_{a}=0$ on the corresponding non-conformal Mellin amplitudes. For example, for the single propagator case, to obtain the scale invariant Mellin amplitude, we would need to impose $\lambda_{1}+\lambda_{2}=0$ on \eqref{offshellpropagator}. 

\vspace*{0.06in} From the examples given in subsection \ref{some_examples}, we can see that each amplitude has a factor involving the product over Gamma functions, namely, $\prod_{i}\Gamma(\rho^{i})$ (we have suppressed the label for the internal vertices). We also know that the conformal Mellin amplitudes involve product over delta functions with the same arguments, namely, $\prod_{i}\delta(\rho^{i})$. In terms of dual Mellin momenta, we can write
\begin{eqnarray}
\rho^{i}=\Delta^{i}-\sum_{j\neq i}s^{ij}=(k^{i})^{2}+\Delta^{i}     \non 
\end{eqnarray}
Thus, in the conformal case, the delta functions impose a set of constraints $(k^{i})^{2}+\Delta^{i}=0$ which is the ``on-shell'' condition for the Mellin momenta. In contrast, for the scale invariant amplitudes, we have Gamma functions with the same arguments for each external leg. In the space of Mellin momenta, the ``on-shell point'' (or equivalently the conformal theory) lies at the pole of the Gamma function (where its argument vanishes). This motivates us to interpret these Gamma functions as external leg factors. It is in this sense that the scale invariant amplitudes are ``off-shell'' objects and imposing the conformality conditions is akin to an LSZ prescription in which the external leg factors are replaced by the corresponding delta functions. It would be very nice to take this observation further and put it on a more rigorous footing. 

\section{Discussion}

In this draft, we have taken some steps in formulating Feynman rules in Mellin space for weakly coupled CFTs. For simplicity, we have restricted to scalar operators.  We considered first the case when the weakly coupled CFT is defined as an exactly marginal perturbation of a free CFT. In this context, we were able to prove in complete generality that the Mellin amplitude for any tree level Feynman diagram involving only the scalar operators factorises into a product of beta functions (with appropriate arguments involving Mellin variables) each of which is associated with a propagator. The meromorphy of the Mellin amplitudes and the identification of its poles with the exchanged primary and its descendants is manifest from this result. We also gave a diagrammatic algorithm to write down the Mellin amplitude of any diagram (tree as well as loops) in terms of integrals over the internal Schwinger parameters. These are the main results of this draft. 

\vspace*{0.06in}{\color{black} Thereafter, we undertook the study of one loop conformal Mellin amplitudes. However, we have not been able, so far, to generalize the tree level Feynman rules to loops. We discussed the scenario where a generic scalar perturbation about a free CFT breaks the conformality of the theory. In particular, even if we add a classically marginal perturbation, the loop corrections will, in general, break the marginality of the interaction rendering the perturbative interacting theory non-conformal. We extended the definition of the Mellin amplitudes as provided by Mack \cite{Mack1} to such a setting where we have only one constraint restricting the number of independent Mellin variables. We calculated some simple examples of these non-conformal Mellin amplitudes. The Beta function conformal propagator uplifts to a $_{3}F_{2}$ hypergeometric function in the non-conformal case. We also considered position space correlators in theories with scaling symmetry  but not the full conformal symmetry. The corresponding Mellin amplitudes seem to be like ``off-shell'' objects which are related to the ``on-shell'' conformal Mellin amplitudes through an LSZ like prescription.} 

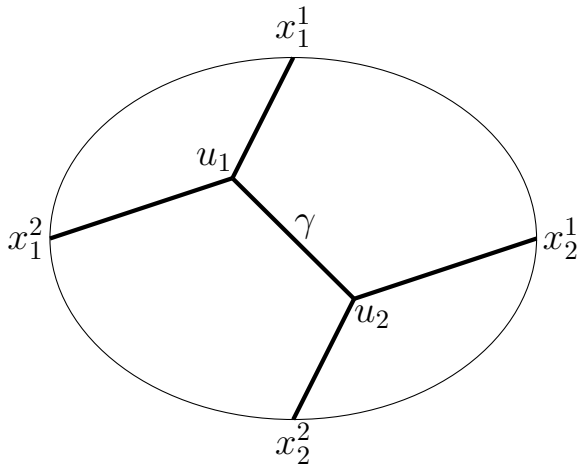
\begin{figure}[h]
\begin{center}
\begin{tikzpicture}[scale=.8]
\draw (0,0) ellipse (4cm and 3cm);
\draw [-] [ultra thick](0,3) -- (-1,1);
\draw [-] [ultra thick](-4,0) -- (-1,1);
\draw [-] [ultra thick](1,-1) -- (-1,1);
\draw [-] [ultra thick](4,0) -- (1,-1);
\draw [-] [ultra thick](0,-3) -- (1,-1);
\draw (-1.3,1.3) node{\Large $u_1$};
\draw (1.3,-1.3) node{\Large $u_2$};
\draw (-4.4,0) node{\Large $x_1^{2}$};
\draw (4.4,0) node{\Large $x^{1}_2$};
\draw (0,-3.4) node{\Large $x_2^{2}$};
\draw (0,3.5) node{\Large $x_1^1$};
\draw (0.2,0.2) node{\Large $\gamma$};
\end{tikzpicture}
\end{center}
\caption{4-point exchange Witten diagram}
\label{diag:strcoup}
\end{figure}

\vspace*{0.06in}One can compare these results at weak coupling to those at strong coupling obtained using Witten diagrams in the dual bulk theory in $AdS$ in \cite{pened, paulos2, suvrat}. As an example, we look at the result obtained in \cite{paulos2} for the 4-point function exchange Witten diagram involving scalars as shown in Figure \ref{diag:strcoup}. 

The Mellin amplitude for this diagram is given by (for a coupling constant $g$), 
\begin{eqnarray}
M(s_{12})&=&\frac{1}{2}\frac{g^{2}}{(s_{12}-\gamma)}\frac{\Gamma\left(\frac{\Delta_{1}^{1}+\Delta_{1}^{2}+\gamma-\frac{D}{2}}{2}\right)\Gamma\left(\frac{\Delta_{2}^{1}+\Delta_{2}^{2}+\gamma-\frac{D}{2}}{2}\right)}{\Gamma\left(1+\gamma-\frac{D}{2}\right)}  \label{dis1}\non \\
 \nonumber && _{3}F_{2}\left(\frac{2-\Delta_{1}^{1}-\Delta_{1}^{2}+\gamma}{2},\frac{2-\Delta_{2}^{1}-\Delta_{2}^{2}+\gamma}{2},\frac{\gamma-s_{12}}{2};\frac{2+\gamma-s_{12}}{2},1+\gamma-\frac{D}{2};1\right)     
\end{eqnarray}
This Mellin amplitude has the same analytic structure as our corresponding result \eqref{twov10} for the weak coupling case which is proportional to $\beta\left(\frac{\gamma-s_{12}}{2},\frac{D}{2}-\gamma\right)$. 

\vspace*{0.06in}It would be interesting to understand the extrapolation of the weak coupling results to the strong coupling results (in the particular example we have considered, how the beta function of the weakly coupled regime extrapolates to the $_{3}F_{2}$ hypergeometric function in the strong coupling regime). In the maximally supersymmetric case, it may be possible to use the integrability of the boundary field theory as well as the string theory in the bulk to understand this interpolation between the results at strong coupling and at weak coupling.

\vspace*{0.06in}Our analysis throughout was somewhat limited as it dealt only with scalar operators. It is important to extend this perturbative formalism to include tensor and spinor operators as well. There are examples of 2 dimensional CFTs which only involve scalar fields\footnote{We would like to thank R. Loganayagam for drawing our attention to this.}. For such theories, one can try to apply this formalism. However, for application to higher dimensional CFTs, the formalism needs to be extended to include spinning operators. It will also be nice if the tree level Feynman rules can be generalized to loop diagrams. One may expect, in analogy with standard momentum space Feynman rules, that the loop amplitudes can be expressed as an integral over undetermined loop variables, with a product of beta functions (with appropriate arguments) in the integrand. If this is the case, it should be possible to systematically develop conformal perturbation theory in Mellin space to arbitrary loop order.

\acknowledgments

We all are greatly indebted to Rajesh Gopakumar for suggesting this project, for his continued help and guidance throughout the work and for his valuable comments on the draft. We would also like to thank R. Loganayagam for useful discussions and comments on the draft. The research of AR leading to these results have received funding from the European Research Council under the European Community's Seventh Framework Programme (FP7/2007-2013) / ERC grant agreement no. [247252]. AR would like to thank TIFR, Mumbai and Shubhadeep Biswas for providing hospitality during this project. For a considerable duration of the project, SS was affiliated to the Indian Institute of Science Education and Research, Pune. SS acknowledges the hospitality and support of Harish Chandra Research Institute, Allahabad where a major part of this work was completed. SS also acknowledges the hospitality and support of the International Center for Theoretical Sciences, Bengaluru. SS is also thankful to Sunil Mukhi for useful discussions and Dhritiman Nandan for useful discussions and comments on the draft. The work of MV was also supported by the SPM research grant of the Council for Scientific and Industrial Research (CSIR), India.

\appendix
\section{Notations and Conventions}
\label{sec:notation}
\subsection*{Notations for Most Used Variables}
\begin{enumerate}
\item Mellin variable $\equiv \mv$
\item Internal Schwinger parameters $ \equiv t$
\item External Schwinger parameters $\equiv \alpha$
\item Conformal dimension of internal lines $\equiv \inc $
\item Conformal dimension of external lines $\equiv \exc  $
\item Coordinates of external vertices $\equiv x$
\item Coordinates of internal vertices $\equiv u$
\item Number of space-time dimensions $\equiv D$
\end{enumerate}

\subsection*{Convention for Indices}
\begin{enumerate}
\item The vertices are labeled by indices $a,b,\cdots$
\item The co-ordinate of the vertices are $u_a,u_b,\cdots$
\item The co-ordinate of all external points attached to the $a^{th}$ vertex are $x^{i}_a$ (we suppress the spacetime Lorentz index)
\item The conformal dimension of the operator inserted at $x_a^i$ is $\exc_a^i$
\item The squared distance between two points $x^{i}_a$ and $x^{j}_b$ is $$(x_a^i-x_b^j)^2\equiv (x^{ij}_{ab})^2= (x^{ji}_{ba})^2$$
\item The Mellin variable dual to $x_{ab}^{ij}$ is denoted by $\mv_{ab}^{ij}$ which satisfies
$$\mv_{ab}^{ij}=\mv_{ba}^{ji}\qquad\mbox{and}\qquad s_{aa}^{ii}\equiv0$$
\end{enumerate}

\subsection*{Convention for Summations and Products}
\begin{enumerate}
\item If there are $N_a$ external lines meeting at the $a^{th}$ vertex, then we denote
\begin{eqnarray}
\sum_{i\in a}&\equiv & \sum\limits_{i=1}^{N_a}= \textrm{ sum over all {\it external} lines connected to the vertex $u_a$}
\nn\\
\prod_{i\in a}&\equiv &\prod\limits_{i=1}^{N_a}= \textrm{ product over all {\it external} lines connected to the vertex $u_a$}\nn
\end{eqnarray}
\item For the double summation and products (which avoid over counting), we use the notations
\begin{eqnarray}
\sum\limits_{{1\le i<j\le N_a}}\sum \equiv\sum_{(i,j)\in a}
\qquad;\qquad
\prod\limits_{{1\le i<j\le N_a}}\prod \equiv\prod_{(i,j)\in a}\nn
\end{eqnarray}
\item If the upper index is {\it not} mentioned then it implies that upper indices have been summed over all possible values. e.g. for the Mellin variables, we shall use
\begin{eqnarray}
\mv_{aa}\equiv\sum\limits_{(i,j)\in a}\mv_{aa}^{ij}\quad;\quad 
\mv_{ab}=\mv_{ba}\equiv\sum\limits_{i\in a}\sum\limits_{j\in b}\mv_{ab}^{ij}\quad,\quad (a\not=b) 
\nonumber
\end{eqnarray}
\end{enumerate}

\subsection*{Some Other Conventions}
\begin{enumerate}
\item Mellin measure  \hspace{130pt} $[d\mv_{ij}]\equiv\frac{d\mv_{ij}}{2\pi { i}} $
\item Position space measure \hspace{92pt} 
$\mathcal{D}u\equiv\frac{d^D u}{2(2\pi)^{-D/2}} $
\item Mellin space delta function \hspace{75pt} $\delta_M(s-s_0)\equiv2 \pi { i} \ \delta(s-s_0)$
\end{enumerate}

\subsection*{Shorthand Notations}
The Schwinger parameters in the integral expression of Mellin amplitude, for $n$-vertex simple tree and one loop diagrams we consider in this draft, appear in a nice structure. It is useful to introduce a short hand notations for these functions of Schwinger parameters. These notations turn out to be especially convenient for various manipulations. Along with these, we shall also introduce some functions of the Mellin variables.

\subsubsection*{Set 1 : $G_a^c$}
\begin{eqnarray}
G_a^c&\equiv&1+t_{a,a+1}^2(1+t_{a+1,a+2}^2(\cdots\cdots+ t_{c-1,c}^2))   \quad \qquad 1\le a\le n-1\qquad 
\nonumber\\
G_a^a&\equiv& 1 \quad \qquad \quad \qquad 1\le a\le n-1
\label{fourv14}
\end{eqnarray}
The value of the upstair index $c$ will be greater than or equal to the lower index. 


\subsubsection*{Set 2 : $\tilde H_a^b$}
\begin{eqnarray*}
\tilde H_a^b&\equiv&t_{a, a+1}\cdots t_{b-1,b} G_b^{n-1} \quad \quad \quad 1\le  a<b\le n-1
\nonumber\\
\tilde H_a^a&\equiv& G_a^{n-1}\qquad \qquad \quad \quad \quad 1\le  a\le n-1
\nonumber\\
\tilde H_a^n&\equiv&0 \qquad \qquad  \quad \quad \quad 1\le  a\le n 
\end{eqnarray*}
$\tilde H_a^b$ is not symmetric in its indices. A good mnemonic worth remembering is that 
 upstairs index is always larger than or equal to the downstairs index.


\subsubsection*{Set 3 : $K_a$ and $\tilde K_a$}
\begin{eqnarray}
 K_a &\equiv&t_{a,a+1}\cdots t_{n-1,n} \quad \quad 1\le a \le n-1
\nonumber
\\ K_n&\equiv&1
\non\\
\tilde K_a &\equiv&\Big(t_{1n} t_{12} \cdots t_{a-1,a} G_a^{n-1}+K_a
\Big)\quad \quad 1\le a \le n-1
\nonumber\\
\tilde K_n&=&1
\non
\end{eqnarray}

\subsubsection*{Set 4}
\begin{eqnarray}
R_a&\equiv& \inc_{a,a+1}-
\sum_{c=a+1}^{n}\sum_{b=1}^{a}\left(\mv_{bc}\right)\qquad; \qquad 1\le a \le n-1
\non
\end{eqnarray}

\section{Mellin Transformation}
\label{sec:Mellint}
The Mellin transformation of a complex valued function $f(x)$ of real variable $x$, is defined as 
\begin{equation}
\mathcal{M}\{f(x)\}\equiv F\left(s\right)=\int_{0}^{\infty}x^{s-1}f\left(x\right)dx
\label{meltr1}
\end{equation}
The complex Mellin variable $\mv$ is restricted to those values for which the above integral is convergent. In general, the Mellin transform of $f(x)$ exists in a vertical strip in the complex $s$ plane (analytic extension is usually possible). 

\vspace*{.06in} The inverse Mellin transformation is given by 
\begin{eqnarray}
f\left(x\right)
&=&\int_{c-\iifty}^{c+\iifty}[d\mv]\ F\left(\mv\right)x^{-\mv}
\non
\label{meltr2}
\end{eqnarray}
where the constant $c$ lies within the vertical strip in which the integral in \eqref{meltr1} converges. 

\vspace*{.06in}One well known example of the Mellin transform is the Gamma function which can be represented as the Mellin transformation of $e^{-x}$ 
\begin{eqnarray}
\Gamma(s)=\int_0^\infty dx\; x^{s-1}e^{-x}
\non
\end{eqnarray}
with the inverse transformation given by Cahen-Mellin integral
\begin{eqnarray}
\exp\bigl({-x}\bigl)=\int_{c-\iifty}^{c+\iifty} [d\mv]\ \Gamma(\mv)\ x^{-\mv} , \ \ \ c >0
\non
\label{mbiden1}
\end{eqnarray}
\subsection{Mellin Space Delta Function}
\label{mellin_delta}
In this section we want to show that 
  \be
  I = \int_{c-i\infty}^{c+i\infty}[ds]f(s)\int_0^\infty dt\ t^{s_0-s-1}=\int_{c-i\infty}^{c+i\infty}[ds]f(s)\left(2\pi i\delta(s-s_o)\right)
  \label{mede}
  \ee
  where $c=\mbox{Re}(s_0)$. 
  
 \vspace*{.06in}The above identity essentially shows that inside the contour integration, the real integral  $\int_0^\infty dt\ t^{s_0-s-1}$ behaves as the delta function as long as the real part of $(s-s_0)$ is zero along the contour. In order to prove our claim, we first perform a change of variable $t = e^x$ to get
\be
I =  \int_{c-i\infty}^{c+i\infty}[ds]f(s)\int_{-\infty}^\infty dx\ e^{(s_0-s)x}\non
\ee
Now, since the argument $(s_0-s)$ of the exponential function is purely imaginary along the contour of integration, the $x$ integral is an integration over an oscillating function. Using the delta function representation
\be
\delta(y)=\frac{1}{2\pi}\int_{-\infty}^{\infty} dp\ e^{iyp}\non
\ee
we obtain 
\be
I =  \int_{c-i\infty}^{c+i\infty}[ds]f(s)\left(2\pi i\delta(s_0-s)\right)=f(s_0)
\non
\ee
This proves the desired result \eqref{mede}.

\section{Some Useful Identities}
\label{sec:useiden}
\begin{enumerate}
\item The following Mellin-Barnes representation turns out to be very useful
\begin{eqnarray}
\frac{1}{(1+z)^a}=
\frac{1}{\Gamma(a)}\int_{-\iifty}^{\iifty}[ds] \ z^{-s} \Gamma(a-s)\Gamma(s)
\label{mbusfull1}
\end{eqnarray}

\vspace*{.04in}\item The first Barnes lemma is
\begin{eqnarray}
\int_{c-\iifty}^{c+\iifty}[ds]\;\beta(a+s,b-s)\beta(c+s,d-s)=\beta(a+d,b+c)
\label{mbintegral1b}
\end{eqnarray}

\vspace*{.04in}\item An useful rearrangement identity involving the product of beta functions is
\begin{eqnarray}
	\beta(a-u,u)\beta(d-u,k)=\beta(d-u,u)\beta(d,k) \qquad;\quad   \mbox{provided} \quad a=d+k
	\label{rearrangementid1}
\end{eqnarray}

\vspace*{.04in}\item The recursive integral form of product of beta functions is
\begin{eqnarray}
\prod_{a=1}^{m-1}\int_0^\infty dt_{a,a+1}\bigl(t_{a,a+1}\bigl)^{N_a-1}
\Big(G_a^{m}\Bigl)^{-L_a}
=\prod_{a=1}^{m-1}\frac{1}{2}\beta\Big(\frac{N_a}{2},L_a+
\frac{N_{a-1}-N_a}{2}\Big)
\label{mbusefull2}
\end{eqnarray}
where $N_0\equiv 0$ and $G_a^m$ is defined in appendix \ref{sec:notation}.

\vspace*{.08in}\item The definition of the Hypergeometric function is
\begin{eqnarray}
\hspace*{-.5in}{_{p}F_{q}}\Bigl(a_1,\cdots, a_{p}\;;\;b_1,\cdots,b_{q}\;;\;x\Bigl)
&=& \frac{\Gamma(b_1)\cdots \Gamma(b_q)}{\Gamma(a_1)\cdots \Gamma(a_p)}\sum\limits_{n=0}^\infty \frac{\Gamma(a_1+n)\cdots \Gamma(a_p+n)}{\Gamma(b_1+n)\cdots \Gamma(b_q+n)}x^n
\label{iden1}
\end{eqnarray}
with $|x|<1$.
\vspace*{.03in}\item The recursive integration formula for hypergeometric functions is
\begin{eqnarray}
&&\hspace*{-.8in}\beta\left(a_{p+1},b_{q+1}-a_{p+1}\right){_{p+1}F_{q+1}}\Bigl(a_1,\cdots, a_{p+1}\;;\;b_1,\cdots,b_{q+1}\;;\;x\Bigl)
\nonumber\\
&&=\int_0^1 dt\; (t)^{a_{p+1}-1}(1-t)^{b_{q+1}-a_{p+1}-1}{_{p}F_{q}}\Bigl(a_1,\cdots, a_{p}\;;\;b_1,\cdots,b_{q}\;;\;tx\Bigl)
\label{iden2}
\end{eqnarray}
\vspace*{.02in}\item Following identities relate two ${_{3}F_2}$ hypergeometric functions with different arguments
\begin{eqnarray}
{_{3}F_2}(a_1,a_2,a_3;b_1,b_2;1)
&=&{_{3}F_2}(a_1,b_2-a_2,b_2-a_3;b_2,b_1+b_2-a_1-a_3;1)
\non\\
&&\times \ \frac{\Gamma(b_1)\Gamma(b_1+b_2-a_1-a_2-a_3)}{\Gamma(b_1-a_1)\Gamma(b_1+b_2-a_2-a_3)}
\label{iden3}
\end{eqnarray}
and,
\begin{eqnarray}
&&\hspace*{-.3in}{_{3}F_2}(a_1,a_2,a_3;b_1,b_2;1)
\nonumber\\
&=&{_{3}F_2}(b_1-a_1,b_2-a_1,b_1+b_2-a_1-a_2-a_3;b_1+b_2-a_1-a_2,b_1+b_2-a_1-a_3;1)
\non\\
&&\times \ \frac{\Gamma(b_1)\Gamma(b_2)\Gamma(b_1+b_2-a_1-a_2-a_3)}{\Gamma(a_1)\Gamma(b_1+b_2-a_1-a_2)\Gamma(b_1+b_2-a_1-a_3)}
\label{iden4}
\end{eqnarray}
\item The Gauss identity is 
\begin{eqnarray}
{_{2}F_1}(a_1,a_2;b_1;1) = \frac{\Gamma(b_1)\Gamma(b_1-a_1-a_2)}{\Gamma(b_1-a_1)\Gamma(b_1-a_2)}
\label{iden5}
\end{eqnarray}
\item An integral representation of ${_{2}F_1}$ is 
\begin{equation}
_{2}F_{1}(a,b;c;z)=\frac{1}{B(b,c-b)}\int_{0}^{1}t^{b-1}(1-t)^{c-b-1}(1-tz)^{-a}dt
\label{iden6}
\end{equation}
\vspace*{.01in}\item Following identity relates two ${_{2}F_1}$ Hypergeometric function with different arguments
\begin{eqnarray}
_{2}F_{1}(a,b;c;z)&=&(1-z)^{-a}{_{2}}F_{1}
\Big(a,c-b;c;\frac{z}{z-1}\Big)
\label{iden7}
\end{eqnarray}
\end{enumerate}


\section{Details of Some Tree Level Calculations}


\subsection{Mellin Amplitude of a Completely General Tree}
\label{general_tree_proof}
In this appendix, we carry out the derivation of Mellin amplitude for an arbitrary tree Feynman diagram and show that the amplitude is given by product of beta functions with one beta function for each internal propagator.

\vspace*{.06in} From the diagrammatic rules described in section \ref{sec: diag_rules}, we know that a general Mellin amplitude takes the form
\begin{equation}
M\left(\{s_{ab}\}\right)=\prod\left[ \int_{0}^{\infty}dt_{ab}\frac{(t_{ab})^{\gamma_{ab}-1}}{\Gamma(\gamma_{ab})}\right]F\Bigl(\{t_{ab}\},\{s_{ab}\}\Bigl)   
 \label{TGappen1}
\end{equation}
The $t_{ab}$ is the Schwinger parameter for the propagator joining the internal vertices $a$ and $b$ and the product is over all the internal propagators of the Feynman diagram. The function $F$ is, in general, an arbitrary function of the Schwinger parameters $t_{ab}$ and the Mellin variables $s_{ab}$.

\vspace*{.07in}As mentioned earlier, the function $F$, for any given graph, depends on the order of integration over the position space vertices. For a straight chain of propagators (as in section \ref{sec:simtree}), the natural choice is to go from one end to the other without any jumps. This results in the simplest expression for $F$. For a tree with branches, this option is absent. We shall, however, prescribe an order that gives an expression for $F$ such that the integration over the Schwinger parameters can be performed easily. 

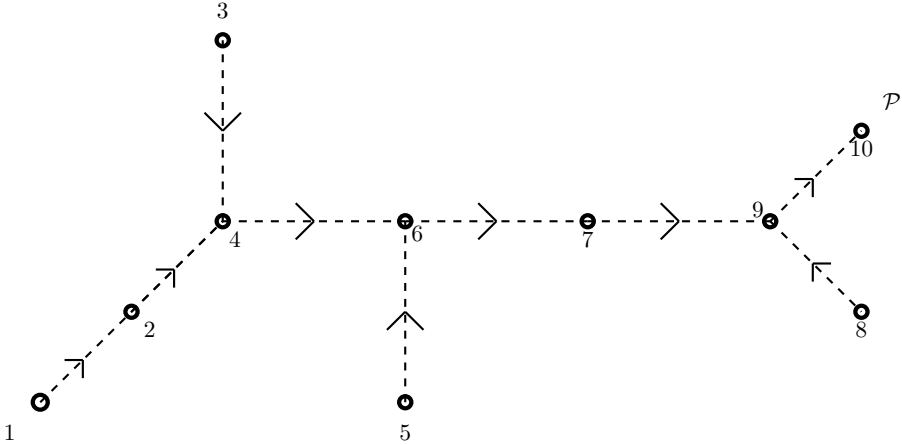
\begin{figure}[h]
\begin{center}
\begin{tikzpicture}[scale=.8,transform shape]
\draw[ultra thick]  (0,0) circle (3.5pt);
\draw[ultra thick]  (3,3) circle (0.10cm);
\draw[ultra thick]  (3,6) circle (0.10cm);
\draw[ultra thick]  (1.5,1.5) circle (0.10cm);
\draw[ultra thick]  (6,3) circle (0.10cm);
\draw[ultra thick]  (6,0) circle (0.10cm);
\draw[ultra thick]  (9,3) circle (0.10cm);
\draw[ultra thick]  (12,3) circle (0.10cm);
\draw[ultra thick]  (13.5,4.5) circle (0.10cm);
\draw[ultra thick]  (13.5,1.5) circle (0.10cm);
\node at (-0.5,-0.5) {$1$};
\node at (1.8,1.2) {$2$};
\node at (3,6.5) {$3$};
\node at (3.2,2.7) {$4$};
\node at (6,-0.5) {$5$};
\node at (6.2,2.8) {$6$};
\node at (9,2.7) {$7$};
\node at (13.5,1.2) {$8$};
\node at (11.8,3.2) {$9$};
\node at (13.5,4.2) {$10$};
\draw[thick, black, dashed] (0,0) -- (3,3);
\draw[thick, black, dashed] (3,3) -- (3,6);
\draw[thick, black, dashed] (3,3) -- (6,3);
\draw[thick, black, dashed] (1.5,1.5) -- (3,3);
\draw[thick, black, dashed] (6,0) -- (6,3);
\draw[thick, black, dashed] (6,3) -- (9,3);
\draw[thick, black, dashed] (12,3) -- (9,3);
\draw[thick, black, dashed] (12,3) -- (13.5,4.5);
\draw[thick, black, dashed] (12,3) -- (13.5,1.5);
\node at (14,5) {$\mathcal{P}$};
\draw[thick, black] (0.7,0.7) -- (0.7,0.4);
\draw[thick, black] (0.7,0.7) -- (0.4,0.7);
\draw[thick, black] (2.2,2.2) -- (2.2,1.9);
\draw[thick, black] (2.2,2.2) -- (1.9,2.2);
\draw[thick, black] (3,4.5) -- (2.7,4.8);
\draw[thick, black] (3,4.5) -- (3.3,4.8);
\draw[thick, black] (4.5,3) -- (4.2,3.3);
\draw[thick, black] (4.5,3) -- (4.2,2.7);
\draw[thick, black] (6,1.5) -- (5.7,1.2);
\draw[thick, black] (6,1.5) -- (6.3,1.2);
\draw[thick, black] (7.5,3) -- (7.2,3.3);
\draw[thick, black] (7.5,3) -- (7.2,2.7);
\draw[thick, black] (10.5,3) -- (10.2,3.3);
\draw[thick, black] (10.5,3) -- (10.2,2.7);
\draw[thick, black] (12.7,2.3) -- (12.7,2);
\draw[thick, black] (12.7,2.3) -- (13,2.3);
\draw[thick, black] (12.7,3.7) -- (12.7,3.4);
\draw[thick, black] (12.7,3.7) -- (12.4,3.7);
\end{tikzpicture}
\end{center}
\caption{Order of integration over the vertices depicted by the numbers (in increasing order)}
\label{vertex_order}
\end{figure}

\vspace*{.07in}To specify the ordering, we first choose any one of the vertices, with only one edge attached, to be the reference vertex $\mathcal{P}$ (see figure \ref{vertex_order}) on the skeleton diagram. In our prescription, the integration over this vertex will be carried out after performing integration over all the other vertices.  For all the other vertices, the order is indicated by some arrows on the lines. For drawing these arrows, one needs to follow two rules. The first rule is that, among all the lines meeting at a vertex, there should be only one line with an outgoing arrow. All the other lines attached to that vertex should have ingoing arrows. The second rule is that any given vertex is integrated only after all the other vertices connected to it by (lines with) ingoing arrows have been integrated over. Thus, $\mathcal{P}$ is the only vertex with a single line which has an ingoing arrow and according to our prescription, it is integrated in the end. Figure \ref{vertex_order} gives an example of a compatible ordering using arrows.  In this example, an order allowed by the above rules is $1 \rightarrow 3\rightarrow 2\rightarrow 4\rightarrow 5\rightarrow 6\rightarrow 7\rightarrow 8\rightarrow 9\rightarrow 10$. This, however, is not the only ordering consistent with our rules and all such allowed orderings are equally good for our purpose.
\vspace*{.06in}

\vspace*{.06in}We can now write down the function $F$ for this prescribed order of integration for any Feynman diagram using the diagrammatic rules described in section \ref{sec: diag_rules}. We state the result here in a graph independent manner. From the diagramatic rules, we know that each pair of vertices $a$, $b$ (which may be the same point also) of the Feynman diagram contributes a multiplicative factor raised to $-s_{ab}$ to $F$. In other words, if we denote this factor by $\mathcal{F}_{ab}$, then the function $F$  is given by,
\begin{equation}
F=\prod_{a=1}^{N}\prod_{b=1}^{N}(\mathcal{F}_{ab})^{-s_{ab}}               
 \label{TGappen2}
\end{equation}  
where $N$ is the total number of internal vertices.

\vspace*{.06in}We now describe the functional dependence of $\mathcal{F}$ on Schwinger parameters. For this, we draw the shortest continuous line (without raising the pen) between the two vertices $a$ and $b$ on the skeleton diagram (via the vertices that come on the way). We refer to the set of vertices that we cross as $\mathcal{A}_{ab}$. Since we are considering a tree, there will exist a vertex in this set that is \textit{nearest} to the reference vertex $\mathcal{P}$. The set of vertices on the continuous route from this vertex to $\mathcal{P}$ is denoted by $\mathcal{B}_{ab}$.

\vspace*{.06in} To see an example, let us consider the pair of points ($3$,$5$) in figure \ref{vertex_order}. Here, $\mathcal{A}_{35}=\{3,4,6,5\}$. The vertex in the set $\mathcal{A}_{35}$ nearest to $\mathcal{P}$ is $6$ and $\mathcal{B}_{35}=\{6,7,9,10\}$. 

\vspace*{.06in}Since the vertices in the sets $\mathcal{A}_{ab}$ and $\mathcal{B}_{ab}$ form individual chains, there is always a line on the skeleton diagram connecting the consecutive vertices in each of these sets. Below, we shall make use of the Schwinger parameters corresponding to the propagators in these lines.

\vspace*{.06in}The functional dependence of $\mathcal{F}_{ab}$ on the Schwinger parameters can now be easily written down by following the diagrammatic rules of section \ref{sec: diag_rules} and is given by
\begin{eqnarray}
\mathcal{F}_{ab}=\left[\prod_{i=1}^{|\mathcal{A}_{ab}|-1}t_{\mathcal{A}_{ab}(i)\mathcal{A}_{ab}(i+1)}\right]K_{ab}    
\label{building_blocks}
\end{eqnarray}
where $|\mathcal{A}_{ab}|$ denotes the number of vertices in the set $\mathcal{A}_{ab}$ and the product in the bracket is over those propagators which lie along the shortest continuous line joining the two vertices $a$ and $b$. The function $K_{ab}$ is given by 
\begin{equation}
K_{ab} \equiv1+t_{\mathcal{B}_{ab}(i)\mathcal{B}_{ab}(i+1)}^{2}K_{\mathcal{B}_{ab}(i+1)\mathcal{B}_{ab}(i+2)}\;\;\;\;\;\;\;\;\;;\;\;\;\;\;\;K_{\mathcal{B}_{ab}\left(|\mathcal{B}_{ab}|-1\right)\mathcal{B}_{ab}\left(|\mathcal{B}_{ab}|\right)}\equiv1     
\non\label{TGappen3}
\end{equation}
 It should be emphasized that the above form of $\mathcal{F}_{ab}$ is true only for the chosen order of integration and will be different if we change the order.

\vspace*{.06in}As an example, for the Feynman diagram in figure \ref{vertex_order}, the factor $\mathcal{F}_{3,5}$ is given by
\begin{eqnarray}
\mathcal{F}_{3,5}=t_{34}t_{46}t_{65}\left(1+t_{67}^{2}\left(1+t_{79}^{2}\left(1+t_{9,10}^{2}\right)\right)\right)
\non
\end{eqnarray}

\vspace*{.06in}By using \eqref{building_blocks} in \eqref{TGappen2} and rearranging the terms, we can write a simplified expression for the function $F$ as   
\begin{eqnarray}
F\;\;=\prod\limits_{\mbox{all propagators}} (t_{ab})^{-P_{ab}}(A_{ab})^{-Q_{ab}}   \label{chap6eqn10}
\end{eqnarray}
where,
\begin{eqnarray}
 &&A_{ab}\equiv1+t_{ab}^{2}\left(1+t_{bc}^{2}\left(1+...(1+t_{op}^{2})...\right)\right)
\non\label{TGadd1}     
\end{eqnarray}
$b$,$c$,...,$o$ are all on the shortest continuous route from $a$ to the reference vertex $\mathcal{P}$. In other words, for each propagator on the skeleton graph, we draw the shortest continuous line connecting it with the propagator containing the reference vertex $\mathcal{P}$. $t_{ab}, t_{bc},\cdots, t_{op}$ are the Schwinger parameters of the successive propagators on this line.

\vspace*{.06in}The functions $P_{ab}$ and $Q_{ab}$ in \eqref{chap6eqn10} are given by

\begin{eqnarray}
\nonumber && P_{ab}=\sum\limits_{c\in L_{ab}}\sum\limits_{d\in R_{ab}}s_{cd} \qquad;\qquad Q_{ab}=\sum_{\substack{c\in L_{ab}\\c\not=a}}\sum_{\substack{d\in \tilde L_{ab}\\d\not=a}}s_{cd}\quad+\sum\limits_{d\in L_{ab}}s_{ad}
 \end{eqnarray}
The $L_{ab}$ ($R_{ab}$) in the above definition refer to the set of vertices which lie to the left (right) of the propagator joining the vertex $a$ and $b$. By convention, we call the set of vertices which include the reference vertex $\mathcal{P}$ as $R_{ab}$. The term involving the double sum in $Q_{ab}$ is absent if not more than two lines meet at the vertex $a$ in the skeleton graph. The tilde in one of the $L$ in this double sum denotes the fact that we should not include terms of the type $s_{cd}$ where $c$ and $d$ are on the same branch in the set $L_{ab}$. 

\vspace*{.06in}Finally, we now need to integrate over the Schwinger parameters in \eqref{TGappen1}. Since the integrals are not factorised, we shall have to carry out one integral at a time and we shall do that in an order compatible with the arrows on the skeleton (integrating over the line involving the reference vertex $\mathcal{P}$ in the end). For example, for the figure ~\ref{vertex_order}, a compatible order is $t_{12}\rightarrow t_{24}\rightarrow t_{34}\rightarrow t_{46}\rightarrow t_{56}\rightarrow t_{67}\rightarrow t_{89}\rightarrow t_{79}\rightarrow t_{9,10}$. Carrying out these integrals in exactly the same way as in the simple tree case in section \ref{sec:simtree} and using the conformality conditions, we obtain the desired result \eqref{chap6eqn24}. 

\subsection{Mellin-Barnes Approach to $n$-Vertex Tree} 
\label{n vertex mellin}
In this appendix, we present an alternative derivation of the factorization of $n$ vertex tree diagram. This derivation makes use of the Barnes' first identity and shows the usefulness of the Barnes integrals for Mellin space. 

\vspace*{.06in}We start by noting that the Mellin amplitude of $n$ vertex simple tree of section \eqref{sec:simtree} can be written in a form which closely resembles the loop amplitude \eqref{eqnloop2}
\begin{eqnarray}
 M_n(\mv_{ab})
&=&
\prod\limits_{a=1}^{n-1}\left[\int_0^\infty dt_{a,a+1}(t_{a,a+1})^{\inc_{a,a+1}-1} \right]
\prod_{b=c}^n\prod_{c=1}^{n-1}\Bigl(\tilde H_c^b+ K_c K_b \Bigl)^{-\mv_{cb}}
\non\label{eqntree3}
\end{eqnarray}
where, the functions $\tilde H_a^b$ and $K_a$ are defined in the appendix \eqref{sec:notation}.

\vspace*{.06in} The basic idea is to use the identity \eqref{mbusfull1} to convert the terms involving sum as products of Mellin Barnes integrals. We then perform the integration over Schwinger parameters. The Mellin Barnes integrations are performed in the end by making repeated use of Barnes first lemma. 

Using the identity \eqref{mbusfull1} and noting that $\tilde H_a^n\equiv 0$ ( which means that we only need to introduce $n(n-1)/2$ Barnes variables $\inmv_{ab}$ ), we obtain after making use of the definitions of $\tilde H_a^b$ and $K_a$ 
\begin{eqnarray}
M(\mv_{ab})&=&
\prod_{b=1}^{n-1}\prod_{c=b}^{n-1}\left(\int_{-\iifty}^{\iifty}[d\inmv_{bc}]\beta(\mv_{bc}-\inmv_{bc},\inmv_{bc})\right)
\int_0^\infty dt_{n-1,n}\bigl(t_{n-1,n}\bigl)^{\lpvr_{n-1}-1}
\nonumber\\
&&
\prod_{a=1}^{n-2}\Bigg(\int_0^\infty dt_{a,a+1}(t_{a,a+1})^{\lpvr_a-1}(G_a^{n-1})^{-\lpvq_a}
\Bigg)
\non\label{eqntree5}
\end{eqnarray}
where,
\begin{eqnarray}
\lpvr_a &\equiv&\trvr_a-2\sum_{b=1}^{a}\inmv_{bb}
-2\sum_{b=1}^{a}\sum_{c=1}^{b-1}\inmv_{bc}\qquad; \qquad 1\le a \le n-1
\nonumber\\
\lpvq_a&\equiv&
\sum_{b=1}^{a}\Big(\mv_{ab}-\inmv_{ab}\Big)\qquad\qquad\qquad\qquad; \qquad 1\le a \le n-1 
\nonumber
\end{eqnarray}
$t_{n-1,n}$ integral is straightforward and it gives a Mellin space delta function. For integration over other Schwinger parameters, we use the identity \eqref{mbusefull2}. After simplifying the expressions by making use of the conformal conditions, we obtain
\begin{eqnarray}
M(\mv_{ab})&=&\prod_{b=1}^{n-1}\left\{\prod_{c=b}^{n-1}
\int_{-\iifty}^{\iifty}[d\inmv_{bc}]\ \beta(\mv_{bc}-\inmv_{bc},\inmv_{bc})\right\}
\prod_{a=1}^{n-2}\frac{1}{2}\beta\Big(\frac{\lpvr_a}{2},\frac{D-2\inc_{a,a+1}}{2}\Big)
\delta_{M}\Big(\lpvr_{n-1}
\Big)
\nonumber\\
\label{eqntreec3}
\end{eqnarray}
where, $\inc_{01}\equiv 0\equiv \lpvr_0$.

\vspace*{.06in}We note that we have obtained beta functions for only $n-2$ propagators. The missing propagator is hidden in the Mellin-Barnes integrals and the Mellin space delta function as we shall show below.

\vspace*{.06in}To perform the integration over the $\inmv_{ab}$ variables, we rewrite \eqref{eqntreec3} as
\begin{eqnarray}
M(\mv_{ab})&=&\prod_{b=1}^{n-2}\left\{\prod_{c=b}^{n-2}
\int_{-\iifty}^{\iifty}[d\inmv_{bc}]\ \beta(\mv_{bc}-\inmv_{bc},\inmv_{bc})\right\}
\prod_{a=1}^{n-2}\frac{1}{2}\beta\Big(\frac{\lpvr_a}{2},\frac{D-2\inc_{a,a+1}}{2}\Big)
\nonumber\\
&&\prod_{b=1}^{n-1}\left\{
\int_{-\iifty}^{\iifty}[d\inmv_{b,n-1}]\ \beta(\mv_{b,n-1}-\inmv_{b,n-1},\inmv_{b,n-1})\right\}
\delta_{M}\Big(\lpvr_{n-1}
\Big)
\label{eqntreespl31}
\end{eqnarray}
The first line does not involve the $\inmv_{a,n-1}$ variables (note that for $n$ vertex case, only $\lpvr_{n-1}$ involves the $\inmv_{a,{n-1}}$ variables). This helps in performing the integration over $\inmv_{a,n-1}$ variables. We first use the delta function to get rid of integration over $\inmv_{n-1,n-1}$ and then use the identity \eqref{mbintegral1b} to perform integration over other $\inmv_{a,n-1}$ variables. After carefully keeping track of various terms, we obtain
\begin{eqnarray}
M(\mv_{ab})
&=&\frac{1}{2}\prod_{b=1}^{n-2}\left\{\prod_{c=b}^{n-2}
\int_{-\iifty}^{\iifty}[d\inmv_{bc}]\ \beta(\mv_{bc}-\inmv_{bc},\inmv_{bc})\right\}
\prod_{a=1}^{n-2}\frac{1}{2}\beta\Big(\frac{\lpvr_a}{2},\frac{D-2\inc_{a,a+1}}{2}\Big)
\nonumber\\
&&\beta\left(\sum\limits_{a=1}^{n-1}\mv_{a,n-1}-\frac{\trvr_{n-1}-\trvr_{n-2}+\lpvr_{n-2}}{2}\;\;,\;\;\frac{\trvr_{n-1}-\trvr_{n-2}+\lpvr_{n-2}}{2}\right)
\label{eqntreespl34}
\end{eqnarray}
The factor of $\frac{1}{2}$ in front arises due to the delta function in \eqref{eqntreespl31}. 

\vspace*{.06in}To proceed further, we note that the conformal condition at vertex 1 can be expressed in following way
 \begin{eqnarray}
\mv_{11}=\frac{\trvr_1}{2}+\frac{D}{2}-\inc_{12}
\non\label{eqntreespl37}
\end{eqnarray}
This allows us to use the rearrangement identity\eqref{rearrangementid1} as
\begin{eqnarray}
\beta\left(\mv_{11}-\inmv_{11},\inmv_{11}\right)\beta\left(\frac{\lpvr_1}{2},\frac{D-2\inc_{12}}{2}\right)=\beta\left(\frac{\trvr_{1}}{2},\frac{D-2\inc_{12}}{2}\right)\beta\left(\frac{\trvr_1}{2}-\inmv_{11},\inmv_{11}\right)
\non
\end{eqnarray}
Using this, we can rewrite \eqref{eqntreespl34} as 
\begin{eqnarray}
M(\mv_{ab})
&=&\frac{1}{2^2}\beta\Big(\frac{\trvr_1}{2},\frac{D-2\inc_{12}}{2}\Big)
\int_{-\iifty}^{\iifty}[d\inmv_{11}]\beta\left(\frac{R_{1}}{2}-\inmv_{11},\inmv_{11}\right)\nonumber\\
&&\int_{-\iifty}^{\iifty}[d\inmv_{12}]\beta\left(\mv_{12}-\inmv_{12},\inmv_{12}\right)
\int_{-\iifty}^{\iifty}[d\inmv_{22}]\beta\left(\mv_{22}-\inmv_{22},\inmv_{22}\right)
\nonumber\\
&&\prod_{b}^{n-2}\left\{\prod_{c}^{n-2}
\int_{-\iifty}^{\iifty}[d\inmv_{bc}]\ \beta(\mv_{bc}-\inmv_{bc},\inmv_{bc})\right\}
\prod_{a=2}^{n-2}\frac{1}{2}\beta\Big(\frac{\lpvr_a}{2},\frac{D-2\inc_{a,a+1}}{2}\Big)
\nonumber\\
&&\beta\left(\sum\limits_{a=1}^{n-1}\mv_{a,n-1}-\frac{R_{n-1}-R_{n-2}+\lpvr_{n-2}}{2}\;\;,\;\;\frac{\trvr_{n-1}-\trvr_{n-2}+\lpvr_{n-2}}{2}\right)
\label{D.15}
\end{eqnarray}
The integrals in the third line do not include the integrations over the $(\inmv_{11},\inmv_{12},\inmv_{22})$ variables. Moreover, these variables appear only in the form of sum ( i.e. $\inmv_{11}+\inmv_{12}+\inmv_{22})$ in the beta functions of last two lines. Hence, it is useful to use a new set of coordinates as follows
\begin{eqnarray}
\{\inmv_{11},\inmv_{12},\inmv_{22}\}\rightarrow \{\inmv_{11},\inmv_{22},u_1\}\qquad;\quad u_1\equiv \inmv_{11}+\inmv_{12}+\inmv_{22}
\nonumber
\end{eqnarray}
After this coordinate change, the last two lines of \eqref{D.15} do not include $\inmv_{12}$ or $\inmv_{22}$ variables anywhere. They just appear in the first two lines. Performing the integration over these variables using Barnes first lemma gives
\begin{eqnarray}
M
&=&\frac{1}{2^3}\beta\Big(\frac{\trvr_1}{2},\frac{D-2\inc_{12}}{2}\Big)
\int_{-\iifty}^{\iifty}[du_1]\beta\left(\frac{\trvr_{1}}{2}+\mv_{12}+\mv_{22}-u_1,u_1\right)\beta\Big(\frac{\trvr_2}{2}-u_1,\frac{D-2\inc_{23}}{2}\Big)
\nonumber\\
&&\prod_{b}^{n-2}\left\{\prod_{c}^{n-2}
\int_{-\iifty}^{\iifty}[d\inmv_{bc}]\ \beta(\mv_{bc}-\inmv_{bc},\inmv_{bc})\right\}
\prod_{a=3}^{n-2}\frac{1}{2}\beta\Big(\frac{\lpvr_a}{2},\frac{D-2\inc_{a,a+1}}{2}\Big)
\nonumber\\
&&\beta\left(\sum\limits_{a=1}^{n-1}\mv_{a,n-1}-\frac{\trvr_{n-1}-\trvr_{n-2}+\lpvr_{n-2}}{2}\;\;,\;\;\frac{\trvr_{n-1}-\trvr_{n-2}+\lpvr_{n-2}}{2}\right)
\non
\end{eqnarray}
In the next step we use the conformal condition at the vertex 2 to use the rearrangement identity \eqref{rearrangementid1} for the two beta functions inside the integration in the first line of the above expression. After this, we note that the $u_1$ variable appears in the combination $u_1+\inmv_{13}+\inmv_{23}+\inmv_{33}$ in all but one of the beta function. We exploit this by trading the $u_1$ variable for a new variable $u_2$ defined as $u_2=u_1+\inmv_{13}+\inmv_{23}+\inmv_{33}$. This allows us to perform the integrations over $\inmv_{13},\inmv_{23}$ and $\inmv_{33}$ variables using the Barnes' first lemma. The end result after this step is
\begin{eqnarray}
M
&=&\frac{1}{2^3}\beta\Big(\frac{\trvr_1}{2},\frac{D-2\inc_{12}}{2}\Big)\beta\Big(\frac{\trvr_2}{2},\frac{D-2\inc_{12}}{2}\Big)
\int_{-\iifty}^{\iifty}[du_2]\beta\left(\frac{\trvr_{2}}{2}+\mv_{13}+\mv_{23}+\mv_{33}-u_2,u_2\right)
\nonumber\\
&&\prod_{b}^{n-2}\left\{\prod_{c}^{n-2}
\int_{-\iifty}^{\iifty}[d\inmv_{bc}]\ \beta(\mv_{bc}-\inmv_{bc},\inmv_{bc})\right\}
\prod_{a=3}^{n-2}\frac{1}{2}\beta\Big(\frac{\lpvr_a}{2},\frac{D-2\inc_{a,a+1}}{2}\Big)
\nonumber\\
&&\beta\left(\sum\limits_{a=1}^{n-1}\mv_{a,n-1}-\frac{\trvr_{n-1}-\trvr_{n-2}+\lpvr_{n-2}}{2}\;\;,\;\;\frac{\trvr_{n-1}-\trvr_{n-2}+\lpvr_{n-2}}{2}\right)\non
\end{eqnarray}
Continuing this process iteratively, i.e. combining the beta functions and then making a change of coordinate (such that only two beta functions involve the appropriate $\inmv_{ab}$ variables), we obtain the desired result. We need to make repeated use of the identity (conformal condition at $a^{th}$ vertex)
\begin{eqnarray}
\frac{\trvr_a}{2}+\sum\limits_{b=1}^{a+1}\mv_{b,a+1}=\frac{\trvr_{a+1}}{2}+\frac{D}{2}-\inc_{a+1,a+2}
\label{conformal_vertex}
\end{eqnarray}
In the end step, we obtain the desired result
\begin{eqnarray}
M(\mv_{ab})
&=&\frac{1}{2}\left[\prod\limits_{a=1}^{n-2}\frac{1}{2}\beta\Big(\frac{\trvr_a}{2},\frac{D-2\inc_{a,a+1}}{2}\Big)\right]\int_{-\iifty}^{\iifty}[du_{n-3}]\beta\Big(\frac{\trvr_{n-2}}{2}-u_{n-3}\;,\;u_{n-3}\Big)
\nonumber\\
&&
\times\ \beta\left(\sum\limits_{a=1}^{n-1}\mv_{a,n-1}-\frac{\trvr_{n-1}}{2}+u_{n-3}\;\;,\;\;\frac{\trvr_{n-1}}{2}-u_{n-3}\right)
\nonumber\\
&=&\prod\limits_{a=1}^{n-1}\frac{1}{2}\beta\Big(\frac{\trvr_a}{2},\frac{D-2\inc_{a,a+1}}{2}\Big)
\non
\end{eqnarray}
where, we have used the identity \eqref{conformal_vertex} for $a=n-1$ after performing the $u_{n-3}$ integration.

\section{Details of Calculation in Section \ref{2vertex_offshell}}
\label{appendix: off_shell}
In this appendix, we present the details of the calculation leading to the equation \eqref{offshellpropagator}. Our starting expression is \eqref{starting_expression}. We insert a partition of unity in this expression in the form
\begin{eqnarray}
1=\int dq_a \;\;\delta\left(q_a-\sum\limits_{i\in a}\Delta_a^i\right) \qquad;\qquad a=1,2
\non
\end{eqnarray}
and make the coordinate transformations $\alpha_a^i=q_a \;y_a^i\;\;( a=1,2)$ and then use the identity \eqref{generalized_beta} to obtain 
\begin{eqnarray}
\tilde M(\mv_{ab})
&=&
\frac{1}{\Gamma(\inc)}\prod\limits_{a=1}^{2}
\left[\frac{\prod\limits_{i\in a}\Gamma(\rho_a^i)}{\Gamma\left(\rho_a\right)}\right]\int_0^\infty dt\;(t)^{R_1-1}\bigl(1+t^2\bigl)^{-s_{11}}\int_0^\infty dq_1 \;(q_1)^{\rho_1-1}
\nonumber\\
&&\int_0^\infty dq_2 \;(q_2)^{\rho_2-1}
\left(q_1\bigl(1+t^2\bigl)+tq_2\right)^{-\lambda_1}
\Bigl(t\ q_1+q_2\Bigl)
^{-\lambda_2}
\nonumber
\end{eqnarray}
where,
\begin{eqnarray}
 R_1\equiv \inc-\mv_{12}\qquad\mbox{and}\quad \rho_a\equiv \sum\limits_{i\in a}\rho_a^i\qquad;\quad a=1,2
\non
\end{eqnarray}
To proceed further, we rescale $q_2\rightarrow tq_1q_2 $ and obtain,
\begin{eqnarray}
\tilde M(\mv_{ab})
&=&
\frac{1}{\Gamma(\inc)}\prod\limits_{a=1}^{2}
\left[\frac{\prod\limits_{i\in a}\Gamma(\rho_a^i)}{\Gamma\left(\rho_a\right)}\right]\int_0^\infty dt\;(t)^{R_1+\rho_2-\lambda_2-1}\bigl(1+t^2\bigl)^{-s_{11}}\int_0^\infty dq_1dq_2
\nonumber\\
&& \;(q_1)^{\rho_1+\rho_2-\lambda_1-\lambda_2-1} \;(q_2)^{\rho_2-1}
\Bigl(1+t^2+t^2q_2\Bigl)^{-\lambda_1}
\Bigl(1+q_2\Bigl)
^{-\lambda_2}
\nonumber
\end{eqnarray}
Integration over $q_1$ gives a delta function. We now take out the factor of $1+t^2$ from the term containing the single power of $\lambda_1$, rescale $t^2\rightarrow t$ and make a further change of coordinates
\begin{eqnarray}
q_2=\frac{u}{1-u}\qquad,\quad t=\frac{v}{1-v}
\non
\end{eqnarray}
This gives,
\begin{eqnarray}
\tilde M(\mv_{ab})
&=&
\frac{1}{2\Gamma(\inc)}\prod\limits_{a=1}^{2}
\left[\frac{\prod\limits_{i\in a}\Gamma(\rho_a^i)}{\Gamma\left(\rho_a\right)}\right]
\delta\Bigl({\rho_1+\rho_2-\lambda_1-\lambda_2}\Bigl)\int_0^1 du\int_0^1dv\;(v)^{\frac{R_1+\rho_2-\lambda_2}{2}-1}
\nonumber\\
&&(u)^{\rho_2-1}\bigl(1-u\bigl)^{\lambda_2-\rho_2-1}
\bigl(1-v\bigl)^{\mv_{11}+\lambda_1-\frac{1}{2}(R_1+\rho_2-\lambda_2)-1}
\Bigl(1+\frac{uv}{1-u}\Bigl)^{-\lambda_1}
\nonumber
\end{eqnarray}
Now, using the identities \eqref{iden6} and \eqref{iden7}, we obtain
\begin{eqnarray}
\tilde M(\mv_{ab})
&=&
\frac{1}{2\Gamma(\inc)}\prod\limits_{a=1}^{2}
\left[\frac{\prod\limits_{i\in a}\Gamma(\rho_a^i)}{\Gamma\left(\rho_a\right)}\right]
\delta\Bigl({\rho_1+\rho_2-\lambda_1-\lambda_2}\Bigl)
\beta\left(\frac{R_1}{2}+\frac{\rho_2-\lambda_2}{2}\;,\;\frac{D}{2}-\inc\right)
\nonumber\\
&&\int_0^1 du\;
(u)^{\rho_2-1}\bigl(1-u\bigl)^{\rho_1-1}
{_{2}F_{1}}\left(\lambda_1\;,\;\frac{D}{2}-\inc\;;\;\mv_{11}+\lambda_1\;;\;u\right)
\nonumber
\end{eqnarray}
We have also used 
\begin{eqnarray}
\mv_{11}+\lambda_1-\frac{1}{2}(R_1+\rho_2-\lambda_2)=\frac{D}{2}-\inc
\non\label{def_k}
\end{eqnarray}

Next, we use the identity \eqref{iden2} for $a=2,b=1$ and the identity \eqref{iden3} to obtain
\begin{eqnarray}
\tilde M(\mv_{ab})
&=&
\frac{1}{2\Gamma(\inc)}\prod\limits_{a=1}^{2}
\left[\frac{\prod\limits_{i\in a}\Gamma(\rho_a^i)}{\Gamma\left(\rho_a\right)}\right]
\beta\left(\frac{R_1}{2}+\frac{\rho_1-\lambda_1}{2}\;,\;\frac{D}{2}-\inc\right)
\beta\Bigl(\rho_1,\rho_2\Bigl)
\nonumber\\
&&{_3F_{2}}\left(\frac{D}{2}-\inc\;,\;\lambda_2\;,\;\rho_1\;;\;\mv_{11}+\rho_1,\lambda_1+\lambda_2\;;\;1\right)\delta\Bigl({\rho_1+\rho_2-\lambda_1-\lambda_2}\Bigl)
\nonumber
\end{eqnarray}
In writing the above expression, we have used the fact that ${_3F_{2}}$ is symmetric in first set of three indices and the second set of two indices. 

\vspace*{.06in}The above expression is not symmetric in the two vertices. We can put it in a symmetric form by using the identity \eqref{iden4}. After some simplification, we obtain the desired result \eqref{offshellpropagator}.


\end{document}